\pdfoutput=1
\RequirePackage{ifpdf}
\ifpdf 
\documentclass[pdftex]{sigma}
\else
\documentclass{sigma}
\fi

\newcommand{\cC}{{\mathcal C}}
\newcommand{\cG}{{\mathcal G}}
\newcommand{\cH}{{\mathcal H}}
\newcommand{\cI}{{\mathcal I}}
\newcommand{\cM}{{\mathcal M}}
\newcommand{\cS}{{\mathcal S}}
\newcommand{\cT}{{\mathcal T}}
\newcommand{\cO}{{\mathcal O}}
\newcommand{\cD}{{\mathcal D}}
\newcommand{\cL}{{\mathcal L}}
\newcommand{\cQ}{{\mathcal Q}}
\newcommand{\SU}{{\rm SU}}

\newcommand\N{{\mathbb N}}
\newcommand\R{{\mathbb R}}

\newcommand{\fc}{\mathfrak{c}}

\newcommand{\Diff}{{\rm Dif\/f}}

\renewcommand\d{\partial}
\newcommand\flow{{\mathbf \alpha}}
\newcommand{\s}{\sigma }
\newcommand{\de}{\delta }

\numberwithin{equation}{section}

\begin{document}

\allowdisplaybreaks

\renewcommand{\thefootnote}{$\star$}

\renewcommand{\PaperNumber}{017}

\FirstPageHeading

\ShortArticleName{Relational Observables in Gravity: a Review}

\ArticleName{Relational Observables in Gravity: a Review\footnote{This
paper is a contribution to the Special Issue ``Loop Quantum Gravity and Cosmology''. The full collection is available at \href{http://www.emis.de/journals/SIGMA/LQGC.html}{http://www.emis.de/journals/SIGMA/LQGC.html}}}

\Author{Johannes TAMBORNINO}

\AuthorNameForHeading{J.~Tambornino}

\Address{Laboratoire de Physique, ENS Lyon, CNRS-UMR 5672, 46 All\'ee d'Italie, Lyon 69007, France}
\Email{\href{mailto:johannes.tambornino@ens-lyon.fr}{johannes.tambornino@ens-lyon.fr}}

\ArticleDates{Received August 31, 2011, in f\/inal form March 14, 2012; Published online March 28, 2012}

\Abstract{We present an overview on relational observables in gravity mainly from a loop quantum gravity perspective. The gauge group of general relativity is the dif\/feomorphism group of the underlying manifold. Consequently, general relativity is a totally constrained theory with vanishing canonical Hamiltonian. This fact, often referred to as the problem of time, provides the main conceptual dif\/f\/iculty towards the construction of gauge-invariant local observables. Nevertheless, within the framework of complete observables, that encode relations between dynamical f\/ields,  progress has been made during the last 20 years. Although analytic control over observables for full gravity is still lacking, perturbative calculations have been performed and within de-parameterizable toy models it was possible for the f\/irst time to construct a full set of gauge invariant observables for a background independent f\/ield theory. We review these developments and comment on their implications for quantum gravity.}


\Classification{83C45; 83C05; 81S05}

\renewcommand{\thefootnote}{\arabic{footnote}}
\setcounter{footnote}{0}

\section{Introduction}
As already noted by Dirac \cite{dirac} the observables of any f\/irst class constraint theory must be constant along the gauge orbits and therefore have vanishing Poisson brackets with all the constraints. In the case of general relativity the gauge group is the four-dimensional dif\/feomorphism group, therefore the theory is fully constrained meaning that the canonical Hamiltonian vani\-shes on the constraint surface \cite{ adm, dirac_gravitation}. Thus, gravitational observables do not evolve dynamically, which is sometimes referred to as the `problem of time' \cite{ProbTime1,isham_prob_time, ProbTime2}.

This nonevolution of observable quantities in gravity is in conf\/lict with our observations as we experience the gravitational interaction as a dynamical process. However, the resolution to this apparent paradox lies in the fact that the canonical Hamiltonian describes \emph{evolution in coordinate time} which, due to general relativity's invariance under general coordinate transformations, is indeed meaningless. What we observe is \emph{evolution with respect to other fields}, for example the matter content of the universe.

From a mathematical point of view it is easy to construct a specif\/ic class of observables, namely spacetime integrals over scalar densities. These quantities can easily be seen to commute with the four-dimensional dif\/feomorphism constraints. However, they do not possess a \emph{local} interpretation, which renders them pretty much useless as far as local observations are concerned. Local observables are much harder to construct: consider for example a scalar f\/ield $\phi(\sigma)$. The action of the spatial dif\/feomorphism constraint on this f\/ield is given by $\d_a \phi(\sigma)$. Unless $\phi(\sigma):= \phi_0$ is chosen constant it cannot be an observable, thus local f\/ields do not qualify as observables for general relativity. In fact, as was noted by Torre \cite{torre}, general relativistic observables have to include an inf\/inite number of derivatives and are therefore very nonlocal.

These problems can partially be resolved in the framework of \emph{complete observables} which, in its present form, was f\/irst introduced by Rovelli \cite{rovelli2, rovelli1}. The main idea is to relate multiple gauge-dependent f\/ields in a gauge-invariant way and thus a precise implementation of the idea that all that matters in general relativity are relations between dynamical quantities. The mathematical underpinning of this formalism was substantially enhanced by Dittrich in \cite{dittrich2, dittrich1} and presented in a form that makes it applicable for full general relativity.

However, applied to general relativity the formalism is rather complicated and therefore calculations could not be pushed beyond the formal level in full analyticity so far. How can local observables, such as for example gravitational waves, be described in a fully gauge-invariant way? Progress into that direction was made in \cite{dittrich_tambo1, dittrich_tambo2} where perturbation theory around a f\/ixed spacetime was developed.

For full general relativity with realistic matter content the  structure of the observable algebra is dif\/f\/icult to obtain analytically. However, it turns out that for certain matter models the theory de-parameterizes and a full set of gauge invariant observables can be derived. This was for example done using pressure-less dust f\/ields \cite{brown_kuchar, ghtw1} or a scalar f\/ield \cite{gravity_quantized, rovelli_smolin_hamiltonian}. The matter f\/ields used there play the role of a dynamically coupled physical reference frame. Disregarding their phenomenological viability these models allowed for a f\/irst time to analytically study the problem of observables in a fully background independent f\/ield theory. Thus, there is hope that the techniques used there can be generalized to realistic matter models such as ordinary standard model matter.

Whereas the question of observables for classical gravity is by now rather well understood, at least on a conceptual level, similar developments for quantum gravity are still lacking. There are candidate theories that provide a framework for quantum dynamics of the gravitational f\/ield at the Planck scale, among them loop quantum gravity \cite{rovellibook, thiemannbook}, but the details of the physical Hilbert space (that is, the quantum analogue of the classical equivalence classes of solutions to Einstein's equations) remain to a large extent unexplored. Therefore, a systematic analysis of quantum gravitational observables could not be carried out so far.

However, there has been a lot of progress using f\/inite dimensional toy models to understand the conceptual dif\/f\/iculties that arise when the principles of quantum mechanics are combined with background independence, most notably by Rovelli and collaborators (see for example~\cite{rovelli_relational_qm}, where a novel (relational) interpretation of quantum mechanics and the wave function is developed). Within these models strategies have been developed to encode dynamics in a~timeless quantum system and it was shown how to re-obtain unitary evolution in some limit.

On the f\/ield theoretic side progress concerning observables for quantum gravity \cite{gravity_quantized, AQGIV} has mainly been made in the de-parameterizable models mentioned above, where the problem of time can be circumvented by choosing a dynamically coupled reference frame. However, as the material reference frame essentially remains classical all the way down to the Planck scale it is still under investigation in how far these models can be trusted in the deep quantum regime.

The f\/ield of observables for classical and quantum gravity is a f\/ield with a long history that still receives a lot of attention. Since the pioneering works of Dirac \cite{dirac_gravitation} and Arnowitt, Deser and Misner \cite{adm}, decomposing spacetime into a one parameter family of spatial slices to cast the theory into Hamiltonian form, it became clear that the question of gauge-invariant (under the dif\/feomorphism group) quantities for gravity is strongly linked to the choice of a reference frame or observer and its physical description in terms of elementary f\/ields. Bergmann and Komar \cite{ProbTime1, bergmann,bergmann_komar, komar} were the f\/irst to systematically analyse this issue and constructed a set of spacetime scalars from gravitational degrees of freedom that serves as a dynamically coupled reference frame. Whereas the interplay between gauge-invariance and reference frames and the `problem of time' are rather well understood at the classical level, there remain many open questions concerning its implications for a quantum theory of gravity: does the concept of `time' in its classical sense even make sense in the deep quantum regime? what is the role of an observer in a theory of quantum gravity? is there any viable probability interpretation? is time evolution unitary in the standard sense? See for example the reviews by Kucha\v{r}~\cite{kuchar_1993} or Isham~\cite{isham_1995} for a~critical acclaim of interpretational problems that have to be overcome by any candidate theory of quantum gravity. Written almost 20 years ago most of the problems presented therein are still unsolved. See also the work of Barbour and collaborators \cite{barbour_machsprinciple, barbour_natureoftime, barbour_bertotti} for a complementary relational view on the problem of time.

In this article we restrict ourselves mainly to the relational approach to the construction of observables for general relativity. We try to give an overview of the relevant developments but, due to the breadth of the f\/ield, such an overview is necessarily selective and represents the author's own view on the f\/ield.

\section{Complete observables for Hamiltonian constraint systems} \label{sec:general}

General relativity is a gauge theory with $\Diff(M)$ as a gauge group, that is, mathematically dif\/ferent solutions to Einstein's equations which are related by a dif\/feomorphism can not be distinguished from physical grounds. Such equivalence classes of solutions are the observables of the theory. In general relativity the gauge structure is strongly interlinked with the question of observers: dif\/ferent observers, related by a dif\/feomorphism, will use dif\/ferent solutions to Einstein's equations to describe the same physical situation. A strategy to compute observables that is especially well suited for this interlink is the method of complete observables, which we discuss in this section.

\subsection{A warmup example} \label{warmup_example}
Before diving into the details of the mathematical formalism we discuss an almost trivial, but nevertheless illuminating, warmup example to illustrate the general idea. Consider a free nonrelativistic particle in one spatial dimension (hence, the conf\/iguration space is one-dimensional)  with an action given by
\begin{gather} \label{action1}
S[q] := \int_{t_1}^{t_2} dt \frac{1}{2}m\dot{q}^2   ,
\end{gather}
where $q$ is the position of the particle, $m$ its mass and a dot indicates a derivative with respect to $t$. Evaluating $S[q]$ on its extremal points gives the solutions $q(\tau) = q + \frac{p}{m}(\tau - t)$, where $p$ denotes the particle momentum. As $S[q]$ has no nontrivial symmetries, there are no constraints and therefore $q(\tau)$ is the time evolution of an observable.  In nonrelativistic physics one usually assumes that the time parameter $t$ corresponds to some external clock which is not inf\/luenced by the dynamics of the system. However, this is not the case in general relativity due to background independence: there is no preferred notion of external time.

To mimic this feature of background independence one can reformulate the nonrelativistic particle and introduce an additional (unphysical) parameter $s$. Consider an action principle on an extended two-dimensional conf\/iguration space
\begin{gather} \label{action2}
S_{\rm ext}[q,t] := \int_{s_1}^{s_2} ds \frac{1}{2} m \frac{{q'}^2}{t'}   ,
\end{gather}
where a slash now denotes derivatives with respect to $s$. One can easily see that both actions generate the same equations of motion because (\ref{action2}) evaluated on paths $(q(s), t(s))$ gives the same value as (\ref{action1}) evaluated on paths $q(s(t))$. Obviously, (\ref{action2}) is invariant under a general re-parameterization $s \rightarrow \tilde{s} := f(s)$ where $f$ is an arbitrary smooth function. This is the analogue of the dif\/feomorphism symmetry in general relativity. Introducing momenta $p_t$ and $p$ associated to the extended conf\/iguration space variables~$t$ and~$q$ respectively one can perform a Le\-gendre transformation to obtain a canonical formulation of that system. However, the Legendre transformation turns out to be singular, which leads to a constraint
\begin{gather} \label{constraint}
\fc := p_t + \frac{p^2}{2m}   ,
\end{gather}
and the Hamiltonian of the extended system is given by
\[
H_{\rm ext} = N(q, p_q, t, p_t) \fc   ,
\]
where $N(q, p_q, t, p_t)$ is an arbitrary function on the extended phase space, analogous to lapse and shift in the canonical formulation of general relativity.

All observables must have vanishing Poisson-brackets with the constraint and, as a consequence, do not evolve in parameter-time $s$. This is analogous to the `problem of time' found in general relativity.

In this toy model there is an easy explanation and solution to this paradox: We introduced the additional parameter-time by hand, therefore evolution in $s$ is meaningless. From~(\ref{action1}) we already know that the only observable is given by $q(\tau)$, that is, it encodes a \emph{relation} between the two dynamical variables~$q$ and~$t$. In more detail the gauge orbits of $q$ and $t$, generated by the constraint $\fc$ are given by
\[
q(s)  :=  \flow^s_{\fc}(q) = q + \frac{p}{m}s, \qquad
t(s)  : =  \flow_{\fc}^s(t) = t + s   .
\]
The latter can trivially be inverted for $s$, and for each f\/ixed value $t(s) \stackrel{!}{=} \tau$ we can insert this into the f\/irst to obtain
\[
q(\tau) = q + \frac{p}{m}(\tau - t)  ,
\]
which can directly be seen to commute with (\ref{constraint}). As we will see soon this is an example of a more general construction principle for gauge-invariant observables. Considering the gauge orbits $\flow^s_{\fc}(q)$ and $\flow_{\fc}^s(t)$ of the two phase space functions $q$ and $t$ an observable is given by
\begin{gather} \label{first_observable}
F^\tau_{q,t}:= \flow_{\fc}^s(q)\big|_{\flow_{\fc}^s(t) = \tau}   ,
\end{gather}
that is, the value of $q$ at that point of the gauge orbit where $t$ takes the value $\tau$. This is an example of the intuitive idea that in a background independent theory all that matters are \emph{relations between dynamical quantities}. In the following we explain how this statement can be made mathematically precise.

\subsection{Partial or complete observables?}

Rovelli \cite{rovellibook} has noted that gauge invariant observables \`a la (\ref{first_observable}) are \emph{not} the only quantities of physical interest in general relativity: his argument is based on the observation that Einstein's equations are applied successfully to numerous astrophysical and cosmological measurements without even bothering about gauge-invariant observables. There one uses \emph{gauge-dependent} \emph{partial observables} (namely the metric, coordinate-distances, etc.) to describe what one observes. A natural resolution of this apparent paradox, from the relational point of view, is the following: the coordinates which one measures are the readings of some physical coordinate system (for example def\/ined by the collection of all matter f\/ields present in the universe). However, typically one does not bother to describe this collection of all matter f\/ields in the universe as it is way to complicated. Thus, observable quantities seem gauge-dependent from the restricted point of view. When including equations for the dynamical evolution of \emph{all} matter into this system one would expect that the partial observables of the restricted system emerge as the complete observables of the full system f\/ixed to some gauge. This could possibly explain the fact that partial observables often seem to describe the correct phenomenology although they are not gauge invariant quantities.

This observation is realized in a precise way in the de-parameterizable matter models which we will discuss further down. Something similar is conjectured to happen when considering a~universe f\/illed with more realistic matter content. There one does not have enough analytic control on the expressions to actually prove that this is truly the case. But the (experimentally well established) fact that one can use Einstein's equations for gauge-dependent partial observables to describe physical observations is a strong indication into this direction.

Many of the results in loop quantum gravity, such as for example the calculation of the graviton propagator from a fully background independent point of view (see \cite{speziale_graviton_review} for a review), are performed on the level of the \emph{kinematical} Hilbert space, thus implicitly using gauge-dependent partial observables. From our point of view (apart from the fact that the gauge used to calculate the propagator is not explicitly known, and consequently also its physical interpretation) such calculations are perfectly f\/ine as they are carried out in the semiclassical regime: by the same argument as for classical gravity one can conjecture the gauge-dependence of \emph{observable} quantities to be related to our ignorance of the surrounding f\/ields that provide a physical reference frame. Taking into account all these f\/ields (which will not be possible practically) would render the quantum partial observables gauge invariant. At least in a regime where a suf\/f\/icient number of degrees of freedom can still be treated classically (to provide the dynamical reference frame) this interpretation will always be possible. Whether a similar scheme holds in the deep quantum regime, where all degrees of freedom need to be treated quantum mechanically, is an open question, as one would encounter all the dif\/f\/iculties that occur within standard quantum mechanics when treating the measurement apparatus quantum mechanically as well. But for the moment we do not consider this too big a problem. Should there be any experimental contact with quantum gravity in the future this will f\/irst concern the semiclassical regime. However, it remains an interesting conceptual question.

\subsection{Detailed construction}
Consider the Hamiltonian formulation of a classical mechanical system with $d$ degrees of freedom\footnote{For simplicity, assume here that $d$ is f\/inite. With more care these methods easily generalize to the f\/ield theoretic case.}:
Let $\cM$ be a $2d$-dimensional symplectic manifold (the phase space)
with symplectic 2-form $\Omega \in \Lambda^2(\cM)$.
We will be interested in the special case where $\cM = T^*\cQ$ has the structure of a~cotangent bundle over some manifold $\cQ$ (the conf\/iguration space) as this is the kind of structure that one gets when starting from an action principle and then performing a Legendre transformation.
To each function $f \in C^\infty(\cM)$ one can associate a \emph{Hamiltonian vector field} as
\[
\chi_f := \Omega^{-1}(df, \cdot)   ,
\]
which in turn def\/ines the Poisson bracket structure on $C^\infty(\cM)$:
\begin{gather*}
 \{ \cdot, \cdot \} :  \ \  C^\infty(\cM) \times C^\infty(\cM) \rightarrow C^\infty(\cM), \qquad
  f,f' \mapsto \{f,f'\} := \Omega(\chi_f, \chi_{f'})   .
\end{gather*}
Let $H \in C^\infty(\cM)$ be a function on phase space and $\chi_H$ the associated Hamiltonian vector f\/ield. Then one can def\/ine the {\it flow} of  $f \in C^\infty(\cM)$ generated by $H$ as a one-parameter family of maps
\[
\flow_H^t: \ \ C^{\infty}(\cM) \rightarrow C^{\infty}(\cM), \qquad f \mapsto \flow_{H}^t(f)   ,
\]
 where the f\/low $\flow_{H}^t(y)$ of a point $y \in \cM$ is the unique solution to the dif\/ferential equation
\begin{gather} \label{flow}
\frac{d}{dt}\flow_{H}^t(y) = \chi_H\big(\flow_{H}^t(y)\big), \qquad \flow_{H}^0(y) = y   ,
\end{gather}
and the action on $C^\infty(\cM)$ is def\/ined via pullback
\[
\flow_{H}^t(f)(y)   :=  f\big(\flow_{H}^t(y)\big)  \qquad \forall\, f \in C^\infty(\cM)   .
\]
Among all possible phase space functions there is one of particular importance: the \emph{Hamiltonian} $H \in C^\infty(\cM)$ which is related to the action functional $S$ of the Lagrangian formalism via a~Legendre transformation. \emph{Dynamical evolution} on phase space is def\/ined via its f\/low: for any phase space function $f$ time evolution is given by
\[
f(t) := \flow_H^t(f)   .
\]
Now assume that there is a set of constraints\footnote{In f\/ield theories one will in general encounter an inf\/inite number of constraints. In this case one has a set $\{ \fc^I \}_{I \in \cI}$ where $\cI$ is an inf\/inite index set.} $\{\fc^I \}_{I \in \cI}$ (for an index set $\cI$) fulf\/illing the f\/irst class property, i.e.\ they form a closed Poisson algebra (possibly with structure functions) $\{\fc^I, \fc^J  \} = f^{IJ}_K \fc^K$. Such f\/irst class constraints typically emerge when the action functional $S$ is invariant under gauge symmetries and, as a consequence, the Legendre transformation is degenerate. The \emph{constraint hypersurface} $\cC$ is def\/ined to be the subspace $\cC \subset \cM$ where all the constraints vanish,~i.e.
\[
\cC := \{ y \in \cM \,|\, \fc^I(y) = 0 ; \; \forall\, I \in \cI \}   ,
\]
and any equality that holds at least on $\cC$ is called a \emph{weak equality}, denoted by $\simeq$:
\[
f \simeq f' \Leftrightarrow f(y) = f'(y)  \qquad \forall\, y \in \cC, \qquad   f,f' \in C^\infty(\cM)   .
\]
Using (\ref{flow}) one can explicitly calculate the f\/low:
\begin{gather}
\flow_{H}^t(f)(y)   =   \sum\limits_{n=0}^{\infty}\frac{t^n}{n!} \{H(y) , f(y)\}_{(n)} ,\label{flow3}
\end{gather}
where $\{H, f\}_{(0)} := f$ and $ \{H, f \}_{(n)} := \{H , \{ H,f\}_{(n-1)} \}$ iteratively. The f\/low fulf\/ills the group property, i.e.\ $\flow_{H}^t(f)\circ\flow_{H}^s(f) = \flow_{H}^{t+s}(f)$.

In this formalism a {\it gauge transformation}
 $\cG$ is a map from the constraint surface to the constraint surface that can be written as a composition of f\/lows generated by a f\/irst class set of constraints  $\{ \fc^I \}_{I \in \cI}$,
\begin{gather}
\cG:  \ \  \cC  \rightarrow \cC, \qquad
  y  \mapsto  \cG(y)  =  \flow_{\sum\limits_{I \in \cI}\beta_I\fc^I}(y) :=\flow_{\sum\limits_{I \in \cI}\beta_I\fc^I}^{1}(y)   , \label{gaugetrafo}
\end{gather}
for arbitrary real parameters $\{\beta_I\}_{I \in \cI}$. If the constraints are algebraically independent, the gauge transformations for each $y \in \cC$ span an $|\cI|$-dimensional submanifold of $\cC$, the {\it gauge orbit}. Gauge transformations on $C^\infty(\cM)$ are then def\/ined via pullback. Note that each gauge orbit remains unchanged if the f\/irst class set of constraints  $\{ \fc^I \}_{I \in \cI}$ is replaced by an arbitrary linear combination thereof. That is, for an arbitrary nonsingular matrix $A^I_J(y)$, the gauge orbit def\/ined by $ \{ \tilde{\fc} \}_{I \in \cI}$ with $\tilde{\fc}^I = A^I_J\fc^J$ is exactly the same.

Physical observables must be gauge invariant, thus are constant along each gauge orbit. Therefore, gauge orbits and not points in phase space are the objects in phase space with an unambiguous physical interpretation. However, one can try to simplify this picture by f\/ixing the gauge and picking out exactly one representative of each gauge orbit: this can be achieved by f\/inding $|\cI|$ algebraically independent phase space functions $T_K(y)$ and f\/ix their values  to $T_K(y) \stackrel{!}{=}\tau_K$. In order to ensure that the gauge f\/ixing singles out one and only one point from each gauge orbit the following to conditions must be fulf\/illed (see e.g.~\cite{hentei}):
\begin{enumerate}\itemsep=0pt
\item The chosen gauge must be accessible. That is, for each point $y \in \cC$ there exists a gauge transformation $\cG$ as def\/ined in (\ref{gaugetrafo}) that maps the point $y$ onto a point $y' \in \cC$ fulf\/illing the gauge-f\/ixing conditions $T_K(y')  \simeq  \tau_K$.
\item The gauge conditions $T_K(y)  \simeq  \tau_K$ must f\/ix the gauge completely, i.e.\ there must be no other gauge  transformation other than the identity left preserving $T_K(y)  \simeq  \tau_K$.
\end{enumerate}
However, for an arbitrary gauge system there exists no constructive scheme that allows to f\/ind such a perfect set of gauge f\/ixing conditions \emph{globally}\footnote{As most investigations in gauge theories are carried out at a perturbative level, for many practical purposes this is not a severe problem: one f\/ixes a gauge which fulf\/ills the criteria stated above in a phase space region larger than the radius of convergence of the perturbation series. However, when interested in nonperturbative properties of a gauge theory, such as conf\/inement of quarks in QCD or nonperturbative calculations in gravity, this is not possible. Therefore it is worth to develop alternative strategies to compute the physical degrees of freedom of a gauge theory. One of them, which f\/its very well with the interlink between gauge symmetry and background independence in general relativity is the method of complete observables. } due to topological obstructions, so called \emph{Gribov ambiguities} (see e.g.~\cite{ticciatibook}).

An alternative approach, which circumvents the need to f\/ix a specif\/ic gauge is the method of \emph{complete observables}, based mainly on ideas of Rovelli\footnote{A dif\/ferent, however related, perspective was developed earlier by Bergmann and Komar, see for example~\cite{bergmann, bergmann_komar, komar}.} (see~\cite{rovelli2, rovelli1} and references therein). The starting point is that all observables are constants along each gauge orbit.  Due to~(\ref{flow3}) this can be written as
\begin{gather*}
\{ \fc^I, O \}   \simeq   0 \qquad \forall\, I \in \cI   . 
\end{gather*}
These quantities are often called \emph{Dirac observables}. For simple systems with one constraint Rovelli found a way to construct Dirac observables that have a very intuitive meaning: let $f, T \in C^\infty(\cM)$ be two arbitrary phase space functions and $\fc \in C^\infty(\cM)$ a single constraint. These phase space functions are referred to as \emph{partial observables}, emphasizing the fact that these still contain gauge degrees of freedom. But one can combine these two in a relational manner to obtain a gauge invariant quantity,
\begin{gather}
F^\tau_{f,T}(y)   :=    \flow_{\fc}^t(f)(y)\big|_{\flow_{\fc}^t(T)(y)=\tau}  . \label{complete}
\end{gather}
These gauge invariant quantities are then called \emph{complete observables}.
Their interpretation  is the following: Let $f$ be a partial observable that associates a value to each point $y$ on the constraint surface $\cC$. The constraint $\fc$ attaches a gauge orbit to each point $y$, given by $\flow_{\fc}^t(y)$,  parameterized by $t$. The same holds for $T(y)$. Now $F^\tau_{f,T}(y)$ associates to each point $y$ on the constraint surface the value  $\flow_{\fc}^{t_0}(f)(y)$  where the specif\/ic value $t_0$ for the parameter $t$ is given by $\flow_{\fc}^{t_0}(T)(y)=\tau$. This construction ensures that the complete observables are constants along the gauge orbit and therefore commute with the constraints.

The properties of the above def\/ined complete observable depend very much on the properties of the partial observables used in this construction. Practically one has to solve the relation $\flow_{\fc}^t(T)(y) = \tau$ for $t$ in order to eliminate $t$ in (\ref{complete}). Generally there is no guaranty that the f\/low of $T$ can be uniquely solved for $t$ and the resulting complete observables (\ref{complete}) will generally be multi-valued. If one requires the complete observables to be unique one must demand that the phase space function $T$ has a  monotonic behavior along the gauge orbit.

The concept of complete observables generalizes to the case with more than one\footnote{This construction includes the case of an inf\/inite number of constraints, needed for the treatment of f\/ield theory. If $\cI$ has uncountable cardinality the sum must be replaced by an appropriate integral in the following.} constraint following  Dittrich \cite{dittrich1}: for an arbitrary  $|\cI|$-dimensional f\/irst class constraint system $\{ \fc^I \}_{I \in \cI}$ a~gauge transformation is given by~(\ref{gaugetrafo}).
Analogous to (\ref{complete}), a complete observable can then be def\/ined as a \emph{relation} between a phase space function $f$ and a set of $|\cI|$ algebraically independent phase space functions~$T_K$, coined \emph{clock variables} in this context:
\begin{gather}
F^{\tau_K}_{f,T_K}(y)   =   \flow_{\sum\limits_{I \in \cI}\beta_I\fc^I}(f)(y)\big|_{ \flow_{\sum\limits_{I \in \cI}\beta_I\fc^I}(T_K)(y)=\tau_K  }   . \label{complete2}
\end{gather}
The interpretation is the the same as before, but the situation gets more complicated because the gauge orbit is not one-dimensional but $|\cI|$-dimensional. In order to get a single-valued complete observable $F^{\tau_K}_{f,T_K}(y)$ all the equations $T_K(y) = \tau_K$ are required to have a unique solution\footnote{Even if there is no unique solution to $T_K(y) = \tau_K$ it is possible to def\/ine the complete observable $F^{\tau_K}_{f,T_K}(y)$ if $\alpha_{\sum\limits_{I \in \cI}\beta_I\fc^I}(f)$ takes the same value for all these solutions. This was formally achieved by introducing ``partially invariant partial observables''  (see~\cite{dittrich1}).}.
One can rewrite (\ref{complete2}) in a more convenient form realizing that the complete observables fulf\/ill a set of dif\/ferential equations:
\begin{gather}
\frac{\partial}{\partial \tau_L}F^{\tau_K}_{f,T_K}(y)  \simeq  F^{\tau_K}_{(A^{-1})_{M}^{L}\{\fc^M,f\},T_K}(y),
\qquad
F^{\tau_K=T_K(y)}_{f,T_K}(y)  =  f(y)  , \label{completedgl}
\end{gather}
where
\begin{gather} \label{integral_kernel}
A_{K}^{J} := \{\fc^J, T_K  \}
\end{gather}
is an $|\cI|\times |\cI|$ matrix\footnote{In the case of an inf\/inite dimensional constraint system $A_K^J(\s, \s')$ is an integral kernel where~$K$,~$J$ are now discrete indices and $\s$, $\s'$ continuous ones. }
 and $(A^{-1})_{K}^J A_{J}^{L}=\delta_{K}^L$ and $ A_{J}^{K} (A^{-1})_K^L = \delta_{J}^{L}$ where summation over repeated indices is assumed. The (formal) solution to (\ref{completedgl}) in terms of an inf\/inite sum is
\begin{gather}
 F^{\tau_K}_{f,T_K}
  \simeq    \sum\limits_{n=0}^{\infty}\frac{1}{n!}\{\tilde{\fc}^{K_n}, \{ \dots, \{ \tilde{\fc}^{K_2}, \{\tilde{\fc}^{K_1} , f\},\dots \}   \nonumber\\
\hphantom{F^{\tau_K}_{f,T_K}
  \simeq    \sum\limits_{n=0}^{\infty}}{}   \times (\tau_{K_1}-T_{K_1})(\tau_{K_2}-T_{K_2})\cdots(\tau_{K_n}-T_{K_n})    , \label{obssum}
\end{gather}
where $\tilde{\fc}^K :=(A^{-1})^K_J \fc^J$.

By computing the Poisson bracket of (\ref{obssum}) with the constraints one can check that it is indeed a weak Dirac observable. There one uses that the set of constraints $\{\tilde{\fc}^I\}_{I \in \cI}$ is ``weakly Abelian'' (see below). Note that~(\ref{obssum}) so far is only a formal expression,  neither there is a~guarantee that~$A_K^J$ can be inverted throughout the whole phase space nor that the inf\/inite sum will eventually converge. These properties will in general depend strongly on the choice of the clocks~$T_K$. However, (\ref{obssum}) serves as the starting point for an illuminating perturbative analysis~\cite{dittrich_tambo1, dittrich_tambo2}, which we discuss in Section~\ref{sec:perturbative}.

The constraints $\tilde{\fc}^K$ def\/ined above describe the same gauge orbits as the original ones $\fc^K$ and furthermore fulf\/ill the following important properties as can easily be calculated~\cite{dittrich1}:
\begin{gather} \label{weakly_abelean}
\big\{ \tilde{\fc}^J, T_K  \big\}  \simeq  \delta^{J}_{K}, \qquad
\big\{\tilde{\fc}^K, \tilde{\fc}^L\big\}  =  \cO\big(\fc^2\big) \simeq 0   ,
\end{gather}
where $\cO(\fc^2)$ denotes terms at least quadratic in the constraints, for which reason the latter property is also referred to as `weakly Abelian' in the literature. Introducing these weakly Abelian constraints instead of the original set was the key point that allowed Dittrich to express the Dirac observables of an arbitrary constraint system as an inf\/inite power series as in~(\ref{obssum}). In Section~\ref{sec:deparameterization} we will see that, for a peculiar class of clock f\/ields, the above weak equalities can be replaced by strong equalities (i.e.\ they hold everywhere in phase space) which leads to further simplif\/ications. For a detailed discussion of the mathematical properties of these complete observables from a symplectic geometric point of view see~\cite{thiemann1}. For example, one can show that associating complete observables to partial observables is an algebra morphism,
\[
F^{\tau_K}_{f\cdot g + h, T_K}   =   F^{\tau_K}_{f, T_K}\cdot F^{\tau_K}_{g , T_K} + F^{\tau_K}_{h , T_K}   .
\]

\section{Complete observables for general relativity} \label{sec:relativity}

The formalism discussed in the last section can in principle be applied to any f\/irst class gauge system. However, the relational approach unfolds its true value when applied to general relativity. There the gauge symmetry (invariance under dif\/feomorphism of the four-dimensional manifold) is directly linked to the choice of observer which can ideally be accounted for within the relational approach. However, due to the complexity of the gauge group analytic results for full general relativity coupled to realistic matter f\/ields have not been developed beyond the formal level so far. Nevertheless, there exist some interesting formal results mainly due to Dittrich~\cite{dittrich2}. Furthermore, perturbation theory has been developed which links the complete observables~(\ref{obssum}), evaluated in the neighborhood of a f\/ixed phase space point of symmetry-reduced sector therein, to the dynamics of standard f\/ield theory on a given background \cite{dittrich_tambo1, dittrich_tambo2}. Besides that, it turns out that the problem of calculating observables simplif\/ies drastically when applying the forma\-lism to general relativity coupled to a specif\/ic class of matter f\/ields\footnote{This observation was f\/irst made by Brown and Kucha\v{r} in~\cite{brown_kuchar}. They are not explicitly using the language of relational observables there but the results are essentially the same.}. These matter f\/ields are rather nonstandard and further research is needed to decide  whether the specif\/ic models considered so far have any phenomenological signif\/icance. However, they provide examples where complete observables can be calculated \emph{exactly} in a f\/ield theoretic context and therefore allow some insights in the gauge structure of general relativity. We discuss these developments in this section.

\subsection{General formalism}
The Hamiltonian form of general relativity\footnote{In what follows we mostly use the ADM-formulation of general relativity. At least as far as the classical theory is concerned the construction of observables follows the same scheme when using Ashtekar's variables. There the constraint algebra has to be supplemented by an additional $\SU(2)$-Gauss constraint. However, as these constraints form an ideal in the full constraint algebra, this does not bring any additional obstructions.} was f\/irst introduced by Arnowitt, Deser and Misner in \cite{adm}. The 4-dimensional manifold $M = \R \times \Sigma$ is split into 3-dimensional spatial slices, the geometry on each slice $\Sigma_t$ is described by a Euclidean metric~$q_{ab}(\s)$ and the canonical momentum~$p^{ab}(\s)$ is related to the extrinsic curvature of these spatial slices as embedded in~$M$.  Thus, the phase space~$\cM$ of general relativity is given by the conjugate pair
\begin{gather} \label{gr_phase_space}
\{ q_{ab}(\s), p^{cd}(\s') \} = \delta(\s, \s') \delta_{(a}^c \delta_{b)}^d   ,
\end{gather}
where the round brackets denote symmetrization in the spatial indices. There are four constraints per spatial point, the (spatial) dif\/feomorphism constraint and the Hamiltonian (or scalar) constraint,
\begin{gather}
\fc_a  := -\frac{2}{\kappa}q_{ac}D_{b}p^{bc}, \qquad
\fc  : =  \frac{1}{\kappa}\frac{1}{\sqrt{\det q}}\left[ q_{ac}q_{bd} -\frac{1}{2}q_{ab}q_{cd}   \right]p^{ab}p^{cd} - \sqrt{\det q }R(q), \label{admconstraints}  \end{gather}
where $D_b$ denotes the covariant derivative, $R(q)$ the 3-dimensional Ricci-scalar and $\kappa$ the gravitational coupling constant.

(Gauge-) dynamics in the ADM-formalism is encoded via the canonical Hamiltonian
\[
H_{\rm can}[N, \vec{N}] = \int d\s N(\s)\fc(\s) + N^a(\s) \fc_a(\s),
\]
which vanishes on the constraint hypersurface. The Lagrange multipliers $N$ and $N^a$ are known as \emph{lapse function} and \emph{shift vector} and encode the normal and tangential directions (to the spatial hypersurfaces $\Sigma_t$) of the dynamical f\/low. That the canonical Hamiltonian of general relativity is totally constrained is a feature shared with all background independent theories.

Def\/ining the smeared constraints as
\[
\overrightarrow{\fc}(\overrightarrow{N}) := \int d\s N^a(\s) \fc_a(\s), \qquad {\fc}({N}) := \int d\s N(\s) \fc(s)  ,
\]
 the constraint algebra, sometimes called the \emph{Dirac} or \emph{hypersurface deformation algebra}, is given~by
\begin{gather}
\{\overrightarrow{\fc}(\overrightarrow{N}), \overrightarrow{\fc}(\overrightarrow{N'})\}   =   \kappa \overrightarrow{\fc}(\cL_{\overrightarrow{N}}\overrightarrow{N'}), \qquad
\{ \overrightarrow{\fc}(\overrightarrow{N}), \fc(N) \}   =   \kappa \fc(\cL_{\overrightarrow{N}}N), \nonumber \\
\{  \fc(N), \fc(N') \}   =   \kappa \overrightarrow{\fc}(q^{-1}(NdN' -N'dN)), \label{diracalgebra}
\end{gather}
where $\cL$ denotes the Lie-derivative. This is an inf\/inite dimensional algebra, and furthermore the Poisson-bracket between two Hamiltonian constraints closes only with a (phase space dependent) structure function instead of a constant. Thus, the gauge structure of general relativity is very complicated compared to more common gauge theories such as $\SU(N)$-Yang--Mills theories. From a mathematical point of view this is the main reason why the problem of observables for general relativity is so hard.

However, in \cite{dittrich2,dittrich1} progress was made into that direction due to the following observation: in the canonical formulation of general relativity it is rather straightforward to construct quantities which are invariant under \emph{spatial} dif\/feomorphism, e.g.\ by considering integrals of spatial scalars with density one over $\Sigma$. The construction of \emph{spacetime diffeomorphism invariant} quantities, i.e.\ functions that additionally have vanishing Poisson brackets with the scalar constraint, is much harder, at least when one is interested in quantities that have a local spacetime interpretation. An interesting question is whether one can use these spatially dif\/feomorphism invariant quantities as a starting point to construct full observables for general relativity. The main obstruction which prevents a straightforward implementation of that idea is the fact that the dif\/feomorphism constraints do not form an ideal of the full Dirac algebra and the scalar constraints are no subalgebra. That is, starting from spatially dif\/feomorphism invariant quantities, the corresponding complete observables with respect to the scalar constraints will generally not be invariant under spatial dif\/feomorphisms.

However, in \cite{dittrich1} it was shown that the complete observable (computed with respect to the full Dirac algebra~(\ref{diracalgebra})) $F^{\tau_K}_{f,T^K}$ for any spatially dif\/feomorphism invariant $f$ is (i)~still weakly spatially dif\/feomorphism invariant and (ii)~does not depend on the choice of clock variables associated to the spatial dif\/feomorphism constraints $\fc_a$, as long as all the clock variables commute with~$\fc_a$. This further simplif\/ies the construction of Dirac observables for general relativity as one can directly start with spatially dif\/feomorphism invariant quantities.

For a specif\/ic class of clock variables, namely functions which behave as \emph{spacetime scalars} under the full spacetime dif\/feomorphism group, further simplif\/ications occur. It turns out that the dynamical evolution of a \emph{spatial scalar} $T(\s)$, which behaves under spatial dif\/feomorphisms~as
\[ 
\{ \fc_a(\s), T(\s')  \} = \delta(\s, \s')\partial_a T(\s)   ,
\]
gives rise to a spacetime scalar if and only if it has ultralocal\footnote{A symplectic structure is called ultralocal simply if it is proportional to the delta distribution.} Poisson brackets also with the Hamiltonian constraints
\[ 
\{ \fc(\s), T(\s')  \} \simeq \delta(\s, \s')   .
\]
Hence, when choosing spacetime scalars as clock variables the integral kernel (\ref{integral_kernel}) becomes ultralocal and the weakly Abelian constraints $\tilde{\fc}^I$ are local quantities, simply linear combinations of the original constraints. One can then show~\cite{dittrich2} that the complete observable $F^{\tau_K}_{\psi, T^K}$, where~$\psi(\s)$  and the clock variables $T^K(\s)$, $K = 0, 1,2,3$ are chosen as spacetime scalars, takes a~particular simple form:
\begin{gather}
F^{\tau_K}_{\psi(\s), T^K}    \simeq    \sum\limits_{n=0}^{\infty}\frac{1}{n!}\{\tilde{\fc}^{K_n}[1], \{ \dots, \{ \tilde{\fc}^{K_2}[1], \{\tilde{\fc}^{K_1}[1] , \psi(\s)\},\dots \} \nonumber\\
\hphantom{F^{\tau_K}_{\psi(\s), T^K}    \simeq    \sum\limits_{n=0}^{\infty}}{}
 \times (\tau_{K_1}-T_{K_1})(\tau_{K_2}-T_{K_2})\cdots(\tau_{K_n}-T_{K_n})    ,\label{obssumsimple}
\end{gather}
where
\[
\tilde{\fc}^{K}[1] := \int d\s \tilde{\fc}^K(\s),\qquad  K = 0,1,2,3
\]
is now a \emph{finite} constraint system. That is, by an appropriate choice of clock variables the complete observables for general relativity can be seen to ef\/fectively depend only on $4$ parameters~$\tau^K$, $K = 0,1,2,3 $ instead of inf\/initely many. When all the spacetime scalars are additionally chosen invariant under spatial dif\/feomorphism the complete observables $F^\tau_{\psi, T^0}$ depend on one single parameter~$\tau$ only, which then encodes the dynamical evolution of the observable as measured by a physical clock~$T^0$.

One might wonder whether the choice of spacetime scalars as a starting point for complete observables for general relativity is to restrictive. However, it turns out that the clock-f\/ields~$T^K$, $K=0,1,2,3$ (chosen as spacetime scalars) can be used to transform any tensor on $M$ into a~spacetime scalar, as long as $\det \partial_\mu T^K \neq 0$. For example, the inverse spacetime metric~$g^{\mu \nu}$ can be lifted to a spacetime scalar as
\begin{gather} \label{lift}
g^{KL} := \big(\partial_\mu T^K\big) \big(\partial_\nu T^L\big) g^{\mu \nu}   .
\end{gather}
The clock-f\/ields $T^K$ can thus be understood as dif\/feomorphisms from the spacetime mani\-fold~$M$ to some `physical' manifold $\cT$ which is in turn def\/ined through the possible values of the clock f\/ields.

\subsection{Specif\/ic matter models as a dynamical reference frame} \label{matter_models}
The properties of the complete observables (\ref{obssum}) or (\ref{obssumsimple}) depend crucially on the choice of clock variables~$T^K$ which play the role of a dynamically coupled reference frame. First, the construction uses the weakly Abelian form of the constraint algebra~(\ref{weakly_abelean}), and in order to construct these constraints one has to invert a possibly inf\/inite dimensional matrix. Second, so far there is no control over the radius of convergence (in phase space) of the inf\/inite sum and the validity of~(\ref{obssum}) has to be checked case by case. In fact it is unlikely that one can f\/ind a set of clock variables which lets~(\ref{obssum}) converge throughout the whole phase space, as this would essentially be the same as f\/inding a perfect gauge f\/ixing for general relativity.

\looseness=-1
Thus, for general relativity with realistic matter content the observables are not under control analytically and one has to rely on perturbation theory (see further below). Besides that, it is interesting to study toy-systems where general relativity is coupled to very special\footnote{We consider these matter models very special because they do not represent any of the known standard model matter f\/ields. It is important to note that such matter models have been chosen for reasons of mathematical convenience and not because they provide any interesting phenomenology.} matter models, where observables can be computed analytically. Such matter models include for example pressure-less dust-f\/ields or scalar f\/ields without potential. The common feature shared by all these matter models is that they fall into the class of de-parameterizable constraint systems. See for example~\cite{thiemann2} for a general discussion which matter models lead to de-parameterized constraint systems and also for an application of these ideas to phantom dust with negative energy density.

\subsubsection{De-parameterizable constraint systems} \label{sec:deparameterization}

A constraint system $\{ \fc^I \}_{I \in \cI}$ is said to be de-parameterizable if one can f\/ind a local coordinate chart on phase space with two mutually commuting sets of canonical pairs, denoted by $(q^a, p_a)$ and $(Q_I, P^I)$ such that in this chart the constraints can be written in the locally equivalent form $\fc^I = P^I + h^I(q,p)$ where $h^I$ depends only on $(q^a, p_a)$. Equivalence has to be understood in the sense that the constraint surface spanned by the two sets of constraints is the same. For such constraint systems the construction of observables simplif\/ies a lot. Unfortunately general relativity does not fall into this class. However, by coupling the above mentioned matter models general relativity can be turned into a de-parameterizable theory. The hope therefore is that by analyzing these de-parameterizable toy models one can learn something about the observable structure for general relativity coupled to standard matter.

The reason why de-parameterizable constraint systems are particularly simple is the following: f\/irst class constraints of the form $\fc^I = P^I  + h^I(q,p)$ necessarily form an \emph{Abelian} constraint algebra due to linearity in the momenta $P^I$. The same holds for the generators $h^I$, they also form an Abelian algebra,
\begin{gather} \label{PB_h}
\big\{ h^I, h^J \big\} = 0   .
\end{gather}
Furthermore, when choosing $Q^I$ as clock variables one realizes that property  (\ref{weakly_abelean}) now holds strongly, not only on the constraint surface:
\[
\big\{  \fc^I, \fc^J \big\} = 0, \qquad   \big\{ \fc^I, Q_J \big\} = \delta_J^I   .
\]
For such systems there is no need to introduce weakly Abelian constraints, therefore one of the obstacles mentioned above, namely inverting the matrix $A_I^J$ is absent. When further restricting oneself to phase space functions $f(q,p)$ which do not depend on the clock variables~$Q_I$ and their momenta~$P^I$, the inf\/inite sum~(\ref{obssum}) takes a much simpler form:
\[
F^{\tau_I}_{f(q,p), Q^I} = \sum\limits_{n=0}^\infty \frac{1}{n!} \{ H_{\tau}  , f \}_{(n)}|_{\tau^K \rightarrow (\tau^K - Q_K)},
\]
where the generator of the multi-f\/ingered time evolution is then given by
\[
H_{\tau} := \sum\limits_K \tau^K h^K(q,p).
\]
Due to (\ref{PB_h}) and the f\/irst class property of the de-parameterized constraints $H_{\tau}$ is gauge invariant and further one f\/inds
\[
\frac{\partial}{\partial \tau^K} F^{\tau_I}_{f(q,p), Q^I}  = \big\{ h^K, F^{\tau_I}_{f(q,p), Q^I} \big\}   .
\]
Thus, $H_{\tau}$ is the \emph{physical Hamiltonian} that generates time evolution of observables in the dynamical reference frame def\/ined by the clocks $Q^I$. Furthermore, because $H_\tau$ does not directly depend on the clock f\/ields $Q^I$ this Hamiltonian is time-independent and generates conservative motion.

\subsubsection{Dust as a dynamical reference system} \label{sec:dust}
To give an explicit example of a matter-model that leads to de-parameterization for gravity in this section we will discuss pressure-less dust f\/ields in some detail. We will mainly discuss the model used by Brown and Kucha\v{r} in \cite{brown_kuchar} and later ref\/ined by Giesel, Hofmann, Thiemann and Winkler in \cite{ghtw1, ghtw2}. Similar models were also discussed for example by Kucha\v{r} and Torre~\cite{kuchar_torre}, Rovelli~\cite{rovelli1} or Husain and Pawlowski~\cite{husain_pawlowski_2, husain_pawlowski_1}.

In \cite{brown_kuchar} Brown and Kucha\v{r} realized that the combined system of pressure-less dust coupled to gravity can be brought into a de-parameterized form. They start from the dust action,
\[
S_{\rm dust} = -\frac{1}{2}\int_{M}d^4 X \sqrt{|\det(g)|}\rho
[g^{\mu \nu} U_\mu U_\nu + 1]  , 
\]
which is coupled to the standard Einstein Hilbert action $S_{\rm EH}$. Here $U \in T^*M$ is a one-form def\/ined as the
dif\/ferential  $U = -dT + W_jdS^j$, $j =1,2,3$, for some scalar f\/ields
$T$, $W_j$, $S^j$. Thus, the dust action
is a functional of the spacetime metric $g_{\mu \nu}$ and eight scalar f\/ields $\rho$, $T$, $W_i$,
$S^i$.

This action can be interpreted as a f\/ield theoretic generalization of the
concept of free massive relativistic particles moving on geodesics of
the gravitational f\/ield created by the entire collection of particles, see \cite{brown_kuchar} for further details.

A canonical analysis (after solving some second class constraints that eliminate $\rho$ and $W_j$ and thereafter reducing the dimension of the phase space) leads to a f\/irst class system with constraints
\[
\fc^{\rm tot}  =   \fc + \fc^{\rm dust}, \qquad
\fc^{\rm tot}_a  =   \fc_a + \fc^{\rm dust}_a, 
\]
where the dust contributions are given by
\[
\fc^{\rm dust}   :=  P\sqrt{1 + \frac{1}{P^2}q_{ab}(\d_a T + P_j \d_a S^j) (\d_b T + P_j \d_b S^j)}, \qquad
\fc^{\rm dust}_a  :=  P\d_a T + P_j \d_a S^j   ,
\]
and the geometry contributions are as in (\ref{admconstraints}). $P$ and $P_j$ are the momenta canonically conjugate to $T$ and $S^j$ respectively\footnote{In \cite{brown_kuchar} $P$ is chosen positive, which leads to a standard positive dust energy density. However, in~\cite{ghtw1} it was argued that the more natural choice from the point of view of relational observables is $P<0$. This means that the dust has negative energy density. However, the physical Hamiltonian that generates time evolution in the dust frame for the gravitational variables  after a gauge reduction then turns out to be positive.}. On the constraint hypersurface the constraints $\fc^{\rm tot}$ and $\fc^{\rm
tot}_a$ can be solved for the momenta $P$ and $P_j$ respectively and
 be written in (partially) de-parameterized form as
\begin{alignat}{3}
& \tilde{\fc}^{\rm tot} = P + h, \qquad&&  h = \sqrt{\fc^2 - q^{ab}\fc_a
\fc_b }, & \label{tot_ham3}\\
& \tilde{\fc}^{\rm tot}_j   =   P_j + h_j, \qquad &&  h_j = \big(S^{-1}\big)^a_j [\fc_a - h\d_aT], &  \label{tot_diffeo3}
\end{alignat}
when assuming that $\d_a S^j$ is non degenerate and its inverse is given by $(S^{-1})^a_j$. The total Hamiltonian constraint is in de-parameterized form now, but not the total spatial dif\/feomorphism constraint because $h_j$ still depends on $T$. Thus (\ref{tot_ham3}) and (\ref{tot_diffeo3}) form a strongly Abelian f\/irst class constraint algebra. Furthermore, $\{ h(\s)  , h(\s')  \} = 0 $ but in general $\{ h(\s) , h_j(\s')  \} \neq 0$,  $\{ h_i(\s) , h_j(\s')  \} \neq 0 $.

The f\/ields $T$ and $S^j$ can then be used as clock-variables to compute a complete set of obser\-vables associated to the gravitational variables~$q_{ab}$ and $p^{ab}$. In \cite{brown_kuchar} the terminology of complete observables was not used yet, but it was already observed that one can use the clock variables~$S^j$ to lift any spatial tensor  on $\Sigma$ to the `dust manifold' $\cS:= S(\Sigma)$, as described in~(\ref{lift}). That this is possible is essentially a consequence of the fact that $T$ and $S^j$ are spacetime scalars.
This can be interpreted as a gauge f\/ixing $T(\s,t)=\tau$, $S^j(\s,t)=x^j$ where $\tau\in \mathbb{R}$
and $x \in \cS$. Let $J(\s):=\det(\partial
S/\partial \s)$ and consider
\begin{gather}
\tilde{q}_{ij}(x)  =   \int_{\Sigma}d^3\s |J(\s)|
\delta\big(x^j-S^j(\s)\big)\big(S^{-1}\big)^a_i \big(S^{-1}\big)^b_j q_{ab}(\s),\nonumber \\
\tilde{p}^{ij}(x)  =  \int_{\Sigma}d^3\s |J(\s)|
\delta\big(x^j-S^j(\s)\big)\frac{ (\d_a S^i) (\d_b S^j) p^{ab}}{J}(\s)   .\label{obs_diffall}
\end{gather}
Remarkably it turns out that these two still form a canonical pair:
\begin{gather*}
\big\{ \tilde{p}^{ij}(x), \tilde{q}_{kl}(x')  \big\} = \kappa
\delta_{(k}^i \delta_{l)}^j \delta(x, x')   .
\end{gather*}
The interpretation of (\ref{obs_diffall}) is obvious: These are the
coordinate transformations of $(q_{ab},p^{ab})$ respectively into the
dynamical coordinate system def\/ined by $\s_x=S^{-1}(x)$.

In \cite{ghtw1} Giesel, Hofmann, Thiemann and Winkler then re-derived and expanded these results using the terminology of complete observables: for each of the gravitational variables $f \in \{q_{ab}, p^{ab}\}$ one can calculate complete observables $F^{(\tau, x^j)}_{f,(T, S^j)}$ in the dynamical reference frame given by the readings $(\tau, x^j)$ of the dust-clocks $T$, $S^j$. These are given by
\begin{gather}
Q_{jk}(\tau,x) :=  F^{(\tau, x^j)}_{q_{ab}, (T, S^j)} :=
\sum\limits_{n=0}^{\infty}\frac{1}{n!}
\big\{\tilde{h}_{\tau},\tilde{q}_{jk}\big\}_{(n)}  ,\nonumber
\\
P^{jk}(\tau,x) :=  F^{(\tau, x^j)}_{p^{ab}, (T, S^j)} :=
\sum\limits_{n=0}^{\infty}\frac{1}{n!}
\big\{\tilde{h}_{\tau},\tilde{p}^{jk}\big\}_{(n)}   ,\label{QaP}
\end{gather}
where
\[
\tilde{h}_{\tau}:=\int_{\cal S}\, d^3x \big(\tau -
\tilde{T}(x)\big)\tilde{h}(x)\qquad{\rm with}\quad
\tilde{h}=h(\tilde{q}_{jk},\tilde{p}^{jk})  .
\]
The expressions (\ref{QaP}) can no longer be described in a compact form, they are
rather complicated functions of $\tilde{q}$, $\tilde{p}$.
However, from the reduced phase space point of view this is not a serious obstruction: the relevant quantities are the Dirac observables in the reduced phase space and the physical Hamiltonian which drives their
time evolution with respect to the chosen clock f\/ields.
Setting
$Q_{ij}(x):=Q_{ij}(\tau=0,x)$,
$P^{ij}(x):=P^{ij}(\tau=0,x)$, one can check through an explicit calculation that
\begin{gather} \label{QP_Poisson}
\{ P^{ij}(x), Q_{kl}(x')  \} = \kappa
\delta_{(k}^i \delta_{l)}^j \delta(x, x')
\end{gather}
still form a canonical pair, all other Poisson brackets are vanishing. Then one can def\/ine
\[
H(\sigma) = \sqrt{C^2 - Q^{ij}C_i C_j}(\sigma)  ,
\]
where
\[
C:=\tilde{\fc}\big(\tilde{q}_{jk}=Q_{jk},\tilde{p}^{jk}=P^{jk}\big),\qquad
C_i:=\tilde{\fc}_i\big(\tilde{q}_{jk}=Q_{jk},\tilde{p}^{jk}=P^{jk}\big),
\]
and $\tilde{\fc}$, $\tilde{\fc}_i$ are the gravitational parts of the constraints, $\fc$, $\fc_a$, expressed in the dust
coordinate system. The physical (i.e.\ nonvanishing) Hamiltonian in the dust frame is then given by
\begin{gather}
{H}_{\rm dust} := \int_{\mathcal{S}} d^3 x\,
H(x)   . \label{h_phys1}
\end{gather}
That is, for all observables $f(Q_{ij}, P^{ij})$ evolution in the dust frame is given by
\[
\frac{d}{d\tau}f = \{ { H}_{\rm dust}, f \}   .
\]
The variation of the dust frame Hamiltonian is given by
\begin{gather*}
\delta {H}_{\rm dust}
  =   \int_{\mathcal{S}} d^3x\left[
\left(\frac{C}{H}\right)\delta C - \left(\frac{Q^{ij}C_{j} }{H}\right)\delta
C_{i} + \frac{1}{2 H}C_{i}C_j Q^{ik}Q^{jl}\delta Q_{ij}\right]  \\
\phantom{\delta {H}_{\rm dust}}{}  =:   \int \limits_{\mathcal{S}} d^3x\left( N\delta C + N^i\delta
C_i + \frac{1}{2} H N^i N^j \delta Q_{ij}\right)   .
\end{gather*}
Thus, the dust frame equations of motion generated by
${H}_{\rm dust}$ are almost equivalent to the ones generated by
the canonical Hamiltonian ${H}_{\rm can}$ with the identif\/ication
\[
q_{ab}(\s) \rightarrow Q_{ij}(x), \qquad  p^{ab}(\s) \rightarrow
P^{ij}(x)
\]
modulo the following important dif\/ferences: f\/irst, lapse
$N$ and shift $N^i$ are not phase space independent functions
as for ${H}_{\rm can}$ where they encode the arbitrariness
of the foliation. Rather they are {\it observable} phase space functions
composed out of the elementary f\/ields $Q_{ij}$, $P^{ij}$ as
\[
N := \frac{C}{H},  \qquad N^i = - \frac{Q^{ij} C_j}{H} .
\]
Second, there is one
additional contribution proportional to the Hamiltonian density
$H(x)$.

The dust frame Hamiltonian has an inf\/inite number of conserved charges,
namely energy and momentum density $H(x)$, $C_j(x)$.
The latter ones generate {\it active} dif\/feomorphisms of the
dust space~$\cal S$, they are to be considered as symmetries of the system rather
than gauge transformations generated by the {\it passive}
dif\/feomorphisms of~$\Sigma$. Likewise, the
former are
related in an
intricate way to time re-parameterization invariance in general relativity.

From a mathematical point of view the analysis started in~\cite{brown_kuchar} and then re-derived in the language of complete observables in~\cite{ghtw1} is indeed very interesting: the observation that general relativity can be turned into a de-parameterizable theory when coupling to pressure-less dust f\/ields allowed for a f\/irst time to analytically solve the problem of Dirac observables in a background independent f\/ield theory. This triggered further progress concerning the reduced phase space quantization approach to quantum gravity (see for example \cite{gravity_quantized, AQGIV, husain_pawlowski_1}) which we will comment on in Section~\ref{quantum}. From a physical and phenomenological point of view the implications of that proposal are less clear: as the dust f\/ields were introduced primarily for reasons of mathematical convenience it remains to be investigated whether the analysis can also be adapted to phenomenologically acceptable matter models such as standard model matter.

In \cite{ghtw1, ghtw2} Giesel, Hofmann, Thiemann and Winkler investigated in detail the phenomenological consequences of the dust model. Starting from the hypothesis that the dust f\/ields are truly realized in nature they carefully worked out the consequences that distinguish this model from pure gravity plus standard matter in the cosmological sector. A priori there is one main obstacle to overcome: instead of the two degrees of freedom per spatial point of standard general relativity (two polarizations for a gravitational wave) the dust model has six degrees of freedom per spatial point. This is a consequence of the fact that passive dif\/feomorphism on~$\Sigma$ are replaced by active dif\/feomorphism on~${\cal S}$, the four degrees of freedom present in the dust f\/ields get re-absorbed by~$Q_{ab}$ and~$P^{ab}$, all their components are physically observable \emph{in the dust frame}. However, dust energy- and momentum-density are conserved charges which essentially lets these extra degrees of freedom not dynamically propagate. This observation was used in~\cite{ghtw1, ghtw2} to argue that the additional (i.e.\ not present in the standard treatment) terms found in f\/irst order cosmological perturbation theory can always be tuned arbitrarily small by an appropriate choice of conserved charges which brings this model into perfect agreement with cosmological observations. A similar conclusion was reached in~\cite{gtt} where the spherically symmetric sector of the dust model was analyzed. However, it is unlikely that the parameters of this model can be tuned in such a way to give negligible contributions to gravitational dynamics in any possible conf\/iguration of the gravitational f\/ield: as the dust f\/ields do form singularities in f\/inite time due to their self-gravitation (see for example~\cite{gtt} where such singularities where analyzed in the spherically symmetric sector) they will most likely not provide good clocks for extreme spacetime regions. This is the physical ref\/lection of the mathematical problem to f\/ind a perfect gauge f\/ixing which is valid everywhere in phase space. Furthermore, there is also the following conf\/lict: in order to make the dust model compatible with phenomenology one would tune the dust energy- and momentum density as small as possible to have no measurable ef\/fect whatsoever. On the other hand, the clocks get worse the less energy they carry, heuristically speaking because they are inf\/luenced too much by the system to observe. Of course this argument holds not only for the dust model but needs to be clarif\/ied whenever additional f\/ields are added to the setup just to serve as clock variables.

\subsubsection{Scalar f\/ields as a dynamical reference system} \label{sec:scalar_fields}
Besides the dust-models described above, scalar f\/ields are also frequently used to provide a~dynamically coupled reference frame. At least the Hamiltonian constraint can be brought into de-parameterized form when a single massless scalar f\/ield without potential is coupled to ge\-ne\-ral relativity. This was f\/irst observed by Rovelli and Smolin in~\cite{rovelli_smolin_hamiltonian} and later studied in detail by Kucha\v{r} and Romano in \cite{kuchar_romano}.

The action for a massless scalar f\/ield without potential is given by
\[
S_\phi = \frac{1}{2} \int_M d^4X \sqrt{|\det g|} g^{\mu \nu} (\partial_\mu \phi) (\partial_\nu \phi)   .
\]
After a Legendre transform the canonical brackets are given by $\{\pi(\s), \phi(\s')  \} = \delta(\s, \s')$, where $\pi$~is the scalar f\/ield momentum. A canonical analysis reveals that the scalar f\/ield contributions to dif\/feomorphism and Hamiltonian constraint are given by
\[
\fc^\phi_a = \pi \partial_a \phi , \qquad \fc^\phi = \frac{1}{2 \sqrt{ \det q }} \pi^2 + \frac{1}{2} \sqrt{\det q} q^{ab} (\partial_a \phi) (\partial_b \phi) .
\]
Because there is no potential term for the scalar f\/ield the Hamiltonian constraint for the coupled system gravity plus scalar f\/ield can be written in de-parameterized form as
\begin{gather} \label{scalar_field_ham}
\tilde{\fc}^{\rm tot} := \pi - h, \qquad h := \sqrt{-\sqrt{\det q } \fc + \sqrt{\det q} \sqrt{\fc^2 - q^{ab} \fc_a \fc_b} }   .
\end{gather}
The constraints $\tilde{\fc}^{\rm tot}$, $\fc^{\rm tot}_a$ of the total system gravity plus scalar f\/ield span the same constraint surface as the original set ${\fc}^{\rm tot}$, $\fc^{\rm tot}_a$. Again, because $\tilde{\fc}^{\rm tot}$ is of de-parameterized form these constraints form an Abelian algebra and furthermore $\{h,h\} = 0$ strongly. Besides that, because~$h$ does not depend on~$\phi$ and~$\pi$ it trivially commutes with the~$\fc^\phi_a$. Therefore, the gauge-invariant (note that $h(\s)$ is of density weight one) physical Hamiltonian that generates time evolution with respect to the scalar f\/ield clock is given by
\[
H_\phi := \int d\s h(\s)   .
\]
As described earlier, when considering already spatially dif\/feomorphism invariant partial observables $f$, the associated complete observables do not depend on the clock f\/ields associated to the dif\/feomorphism constraints. Using $\phi$ as the physical clock associated to the Hamiltonian constraint to measure `time' such complete observables are then given by (for spacetime scalars)
\[
F^\tau_{f,\phi} := \sum\limits_{n=0}^{\infty}\frac{\tau^n}{n!} \{ H_\phi , f \}_{(n)}   ,
\]
and therefore
\[
\frac{\partial}{\partial \tau} F^\tau_{f,\phi} = \{ H_\phi , f \}   .
\]
For partial observables which are not yet invariant under spatial dif\/feomorphisms the dynamical evolution is more complicated. Therefore, this model is mainly applicable when the spatial dif\/feomorphism constraint has been solved already by other means.  Recently Domagala, Giesel, Kaminski and Lewandowski~\cite{gravity_quantized} used this model as a starting point for a reduced phase space quantization where the dif\/feomorphism constraints are solved by standard methods on the kinematical Hilbert space of LQG and then dynamics with respect to this reduced Hamiltonian is implemented. This sheds some new light on the link between LQG and loop quantum cosmology where a scalar is also used commonly as a dynamical time variable, compare also~\cite{kaminski_time, kaminski_group_averaging} for an analysis of related issues.

\subsection{Perturbative complete observables for general relativity} \label{sec:perturbative}
For pure general relativity or general relativity coupled to standard matter the strategy described above does not work, essentially because it is not possible to rewrite the constraints such that they are linear in some momentum. For such generic conf\/igurations there is no scheme known so far that allows to compute gauge invariant observables analytically. However, in \cite{dittrich_tambo1, dittrich_tambo2} a~scheme was proposed to compute complete observables \emph{approximately} as perturbations around a f\/ixed point in phase space (such as e.g.~Minkowski space) or around a symmetry-reduced sector in phase space (e.g.\ the homogeneous-isotropic sector that describes cosmology). Within this scheme it can be seen  (for an appropriate choice of clock variables) that the standard observables for f\/ield theory on a given background emerge as zeroth order terms of a perturbative expansion of the full background independent observables. Higher order terms include interactions with gravitational degrees of freedom, namely (classical) gravitons. Thus, the complete observables of full background independent gravity can be understood as a gauge invariant extension of observables of the linearized theory to higher orders.

Let us brief\/ly describe the main strategy in the simpler case of perturbations around a f\/ixed phase space point. The more general case of perturbations around a symmetry-reduced sector is discussed in~\cite{dittrich_tambo2}.

The idea is essentially to expand the complete observable (\ref{obssum}) around a f\/ixed phase space point $Y_0$: let $\cM$ be a $2d$-dimensional phase space with canonical coordinates $\{y^a \}_{a=1}^{2d}$ and  $\{ \fc^I \}_{I \in \cI}$ a f\/irst class constraint system with $|\cI|$ algebraically independent constraints as in Section~\ref{sec:general}. Then choose a phase space point $Y_0$ on the constraint surface and def\/ine f\/luctuations around that phase space point as $\de y^a := y^a - Y^a_0$. Now simply expand all the constraints around $Y^a_0$,
\[
\fc^I = {}^{(0)}\fc^I(Y_0) + {}^{(1)}\fc^I(Y_0, \de y)+ {}^{(2)}\fc^I(Y_0, \de y) + \cdots   .
\]
Here $^{(0)}\fc^I$ depends only on the background variables $Y^a_0$, $^{(1)}\fc^I$ is linear in the f\/luctuations $\de y^a$, $^{(2)}\fc^I$ is quadratic in the f\/luctuations and so on. In general this expansion will have inf\/initely many terms if the constraints are nonpolynomial. Furthermore, the linearized constraints $^{(1)}\fc^I$ fulf\/ill an Abelian constraint algebra,
\[
\big\{ {}^{(1)}\fc^I  , {}^{(1)}\fc^J \big\} = 0   .
\]
Then, starting from (\ref{obssum}) one can def\/ine an \emph{approximate complete observable} $^{[q]}F^{\tau_K}_{f,T^K}$ by simply restricting the full observable to terms that are at most of order $q$ in the f\/luctuations $\de q$. These truncated complete observables do weakly commute with all the constraints $\fc^I$ up to terms of order $q$ in the f\/luctuation. In this sense $^{[q]}F^{\tau_K}_{f,T^K}$ def\/ine observables which are `almost gauge invariant' in a precise manner and whose invariance properties get worse the farther one moves away from $Y_0$ in phase space. There is one technical obstruction though: the full complete observables (\ref{obssum}) are def\/ined using the weakly Abelian constraints (\ref{weakly_abelean}). In order to compute these and to decide which terms are of order $q$ one has to invert the matrix $A_I^K = \{ \fc_I, T^J  \}$ for a given set of clock variables. In order to make $A_I^J$ perturbatively invertible the clocks have to be chosen such that the zeroth order of $A_I^J$ is invertible. Further simplif\/ications occur when the clocks are chosen such that they are canonically conjugate to the linearized constraints. This is as far as general considerations can be carried out.

In \cite{dittrich_tambo1} these ideas were applied to vacuum general relativity (in the Ashtekar formulation) and general relativity  coupled to a massive scalar f\/ield perturbed around Minkowski space: pure vacuum general relativity has two physical degrees of freedom per spatial point. Therefore the ADM-parameterization of this phase space in terms of 3-metrics and 3-momenta as in~(\ref{gr_phase_space}) has a lot of redundant gauge degrees of freedom. In the linearized theory (linearized around Minkowski space) it is well known that the `physical' (i.e.\ having vanishing Poisson brackets with the linearized constraints) are the traceless-transverse modes of the 3-metric (see for example the appendix of~\cite{dittrich_tambo1} for an explanation of this mode decomposition). Thus, one can conveniently use combinations of the `unphysical' (from a linearized point of view) gravitational modes as clock variables. It turns out that in order to make the zeroth and f\/irst order of the complete observables agree with the observables of the linearized theory one can use a generalization of the so-called \emph{ADM clocks} $T_{\rm ADM}^K$~\cite{adm}, certain rather nonlocal expressions involving all gravitational modes except the traceless-transverse ones which are canonically conjugate to the linearized constraints. Unlike scalar f\/ields or dust f\/ields discussed before these clock variables do not have a local interpretation, rather it turns out that they are related to translations at inf\/inity for asymptotically f\/lat spacetimes. Here we will not give the exact form of these clocks as functions of gravitational modes. Neither will we present the rather ugly computations that need to be performed to compute the approximate observables in that dynamical reference frame. All the details can be found in~\cite{dittrich_tambo1}. The main results are the following:
for the case of a massive, minimally coupled scalar f\/ield of mass~$m$ the total constraints are given by
\begin{gather*}
\fc_a^{\rm tot}   :=   \fc^a + \pi \d_a \phi,\qquad
\fc^{\rm tot}   :=   \fc + \frac{1}{2}\left(\frac{1}{\sqrt{\det q}}\pi^2  \sqrt{\det q}q^{ab}\d_a\phi \d_b \phi + \sqrt{\det q} m^2 \phi^2 \right)   ,
\end{gather*}
where the gravitational contributions are as in (\ref{admconstraints}). Interested in the approximative complete observables $^{[q]}F^{\tau_K}_{\phi, T_{\rm ADM}^K}$ associated to the scalar f\/ield in the dynamical reference frame def\/ined by setting the ADM clocks to  $T^0_{\rm ADM} = \tau$, $T^a_{\rm ADM} = 0 $  for low $q$ one f\/inds the following.
\begin{itemize}\itemsep=0pt
\item The zeroth order $^{[0]}F^{\tau,0}_{\phi(\s), T_{\rm ADM}^K} = 0$ because the scalar f\/ield can be interpreted as a~f\/luc\-tua\-tion around the f\/ixed value $\Phi_0 = 0$.
\item The f\/irst order $^{[1]}F^{\tau, 0}_{\phi(\s), T_{\rm ADM}^K}$ gives back the standard dynamical propagation of a free massive scalar f\/ield on a f\/ixed Minkowski background:
\begin{gather*}
^{[1]}F^{\tau,0}_{\phi(\s), T_{\rm ADM}^K} = \int d\s' \cS_\pi(\tau,\s; 0, \s')\pi(\s') + \cS_\phi(\tau,\s; 0, \s')\phi(\s') ,
\end{gather*}
where the propagators are explicitly given by
\begin{gather*}
\cS_\pi(\tau,\s; 0,\s')    :=   \cos\Big[\sqrt{-\Delta_\s + m^2}\tau\Big]\delta(\s, \s),   \\
\cS_\phi(\tau, \s; 0,\s')  :=   \frac{1}{\sqrt{-\Delta_\s + m^2}}\cos\Big[\sqrt{-\Delta_\s + m^2}\tau\Big]\delta(\s, \s)  .
\end{gather*}
\item The second order approximative complete observable $^{[2]}F^{\tau,0}_{\phi(\s), T_{\rm ADM}^K}$ was also computed explicitly in~\cite{dittrich_tambo1} and can schematically be written in a similar form
\begin{gather*}
^{[2]}F^{\tau,0}_{\phi(\s), T_{\rm ADM}^K} = \int\! d\s' \cD^{\rm grav}_\pi(\tau,\s; 0, \s')\pi(\s') + \cD^{\rm grav}_\phi(\tau,\s; 0, \s')\phi(\s') + \mbox{gauge inv. ext.},
\end{gather*}
where the propagators $\cD^{\rm grav}_\pi(\tau,\s; 0, \s')\pi(\s')$ and $\cD^{\rm grav}_\phi(\tau,\s; 0, \s')$ are modif\/ied such as to take interactions with the gravitational f\/ield into account and the gauge invariant extension term vanishes when all but the trace-free-transverse modes are set to zero. Thus, the second order complete observable can be interpreted as a free massive scalar f\/ield propagating on a (nonself-interacting) graviton background. Higher order corrections will also include interaction terms between the gravitational modes themselves.
\end{itemize}

\looseness=-1
The perturbative approach for complete observables was also used in \cite{dittrich_tambo1} to analyze the `spacetime algebra' of observables, that is their Poisson structure. Not surprisingly it was found that the Poisson brackets between two observables, evaluated at dif\/ferent physical times, depends drastically on the clocks used to def\/ine a dynamical reference frame. When computing the Poisson bracket between two scalar f\/ield observables to second order in the reference frame given by the ADM-clocks it turns out that the result is very similar to the f\/lat spacetime case, namely
\begin{gather} \label{spacetime_algebra1}
\Big\{ ^{[2]}F^{\tau_1,0}_{\phi(\s), T_{\rm ADM}^K} , ^{[2]}F^{\tau_2 = 0,0}_{\phi(\s), T_{\rm ADM}^K} \Big\} \propto - \cD^{\rm grav}_\pi(\tau_1,\s; 0, \s') + \cO(2) ,
\end{gather}
where $\cO(2)$ refers to terms of second order. Compared to the f\/lat spacetime case the propagator is replaced by an `ef\/fective' propagator that takes into account to lowest order the ef\/fects of the nontrivial graviton background. When using scalar f\/ield clocks instead of the ADM-clocks as a dynamical reference frame one observes that  (\ref{spacetime_algebra1}) gets correction terms (additionally to the obvious `translation' factor due to the use of a dif\/ferent set of clocks). Interestingly these correction terms are essentially of the form `energy of the observed f\/ield divided by the energy of the clock f\/ields'. Thus, in order to make these correction terms as small as possible one would increase the energy of the clock f\/ields. This ref\/lects the intuition that a physical clock is the better the more energy it has, simply because this way it is less inf\/luenced by the observed f\/ields. However, this is in conf\/lict with a dif\/ferent observation: as in general relativity any local energy density leads to local spacetime curvature one cannot use clock f\/ields with arbitrarily high energies to describe almost Minkowskian spacetime conf\/igurations. Using the ADM-clocks which are rather nonlocal quantities this conf\/lict is somehow circumvented. We will not go into the full f\/ield-theoretic description performed in \cite{dittrich_tambo1} but illustrate this problem using a very simple toy example.

Consider two parameterized nonrelativistic, massless particles described by a constraint
\[
\fc := p_t + \frac{p_1^2}{2m_1} + \frac{p_2^2}{2m_2}  ,
\]
where $p_t$ is the momentum conjugate to the time variable $t$ and $p_1$, $p_2$ are the particle momenta conjugate to their respective position variables $q_1$, $q_2$. As already discussed in Section~\ref{warmup_example} a~complete observable for this system is given by
\begin{gather} \label{observable1}
F^\tau_{q_1,t} = q_1 + \frac{p_1}{m_1}(\tau -t)  .
\end{gather}
This encodes the position $q_1$ of the f\/irst particle when the clock $t$ takes a value $\tau$. The Poisson bracket between two such observables at dif\/ferent times $\tau_1$ and $\tau_2$, analogous to the `spacetime algebra' in the f\/ield theoretic case, is given by
\begin{gather} \label{t_clock_PB}
\big\{ F^{\tau_1}_{q_1,t}  , F^{\tau_2}_{q_1,t}   \big\} = \frac{1}{m_1}(\tau_2 - \tau_1)   .
\end{gather}
Because this is a relational system without any preferred time variable one could also choose a~dif\/ferent clock, namely~$q_2$. The corresponding complete observable is then given by
\begin{gather} \label{observable2}
F^{\tilde{\tau}}_{q_1,t} = q_1 + \frac{p_1}{m_1}\frac{m_2}{p_2}(\tilde{\tau} -q_2)   ,
\end{gather}
and measures the position $q_1$ of the f\/irst particle when the clock $q_2$ takes a value $\tilde{\tau}$. Ignoring the interpretational dif\/ference between $\tau$ and $\tilde{\tau}$ this looks quite dif\/ferent. However, taking into account that the value $\tau$ is reached at that moment at which $\tilde{\tau} = F^\tau_{q_2, t} = q_2 + \frac{p_2}{m_2}(\tau - t)$ and replaces~$\tilde{\tau}$ in~(\ref{observable2}) one gets back~(\ref{observable1}). In so far, both complete observables describe the same evolution as long as the translation between dif\/ferent clocks is taken into account. However, this is not the case for the spacetime algebra:
\[
\big\{ F^{\tilde{\tau}_1}_{q_1,q_2}  , F^{\tilde{\tau}_2}_{q_1,q_2}   \big\} = \left[\frac{1}{m_1}(\tilde{\tau}_2 - \tilde{\tau}_1)\right]
\left[\frac{m_2}{p_2}\right]\left[1 + \frac{p_1^2}{2 m_1}\frac{2 m_2}{p_2^2}\right]   .
\]
The f\/irst factor is equal to the spacetime algebra of the complete observables with respect to the $t$-clock~(\ref{t_clock_PB}) and the second factor is due to the translation between $\tau \rightarrow \tilde{\tau}$. However, the third factor is nontrivial: there is a correction that occurs due to a nonstandard choice of clock that can be recognized as the energy of the observed particle divided by the energy of the clock. Intuitively this correction can be made arbitrary small by increasing the energy of the clock variable~$q_2$. However, in a general relativistic setting this is problematic as discussed before. Thus, there seems to be a fundamental uncertainty when considering the spacetime algebra of observables already at the classical level. This might be related to the quantum uncertainties of relational observables derived by Giddings, Hartle and Marolf in~\cite{giddings_marolf_hartle}.

The implications of this fundamental uncertainty for observables in full general relativity are still rather vague, as the computations performed in~\cite{dittrich_tambo1} are still on a preliminary level. However, the perturbative formalisms for complete observables might provide the right tools to further investigate such fundamental issues. In~\cite{dittrich_tambo2} this formalism was substantially expanded to accommodate for perturbations around a whole symmetry reduced sector in phase space, and as an application f\/irst order cosmological perturbation theory was re-derived from a fully background independent point of view.

An independent perturbative expansion for observables in classical and quantum gravity, in terms of inverse powers of a large cosmological constant, was studied by Gambini and Pullin in~\cite{gambini_pullin_observables_1, gambini_pullin_observables_2}.

\section{Observables for quantum general relativity} \label{quantum}

As discussed in the previous section the situation concerning observables for classical general relativity is well understood, at least from a conceptual point of view. The method of complete observables implements the intuitive idea that observables in a background independent theory should encode \emph{relations between dynamical quantities}. However, the situation of observables for quantum gravity is less clear which, to a large extent, is due to the fact that a generally agreed upon theory of quantum gravity does not exist yet. While loop gravity provides a kinematical framework for the quantum gravitational degrees of freedom and there exist well-motivated proposals for its dynamics, opinions dif\/fer in which exact way this framework is related to classical general relativity. Taking a conservative approach there are mainly two strategies concerning the problem of observables for quantum gravity: the Dirac approach or the reduced phase space approach. In the f\/irst one quantizes the whole phase space (resulting in the kinematical Hilbert space~$\cH_{\rm kin}$) and then constructs quantum versions of the dif\/feomorphism and Hamiltonian constraint. The physical Hilbert space~$\cH_{\rm phys}$, which contains states capturing the gauge invariant information of quantum gravity, is then essentially the joint kernel of all constraint operators. In the latter one  f\/irst computes a complete set of gauge invariant observables of classical general relativity and subsequently only represent these on a Hilbert space. As one quantizes only gauge invariant degrees of freedom there is no need for any quantum constraint operators anymore, one directly constructs the physical Hilbert space. Besides these approaches, there exists also the point of view (as is argued for for example in~\cite{rovelli_25y}) that one should not strictly follow any `quantization recipe' to construct quantum from classical gravity but rather propose the details of the theory by physical intuition and analogy with other quantum f\/ield theories  and then check whether general relativity is contained in its semiclassical limit. Promisingly all these approaches seem to converge to roughly the same picture of quantum spacetime at the Planck scale~\cite{rovellibook, thiemannbook}. However, depending one one's point of view concerning the `derivation' of quantum gravity the problem of quantum observables takes slightly dif\/ferent facettes. Within the reduced phase space approach the observable algebra is typically too complicated to be representable on some Hilbert space, therefore progress into this direction has been limited to the de-parameterizable models discussed above. Within the Dirac approach the physical Hilbert space is still under poor control due to its mathematical complexity which so far prevented progress concerning observables of the full theory. However, there are interesting toy models that allow to study the peculiarities of background independent quantum observables without diving into the details of full loop gravity. We will comment on both approaches below.
Besides that, there has been progress to def\/ine quantum observables directly without constructing them from classical observables.  In \cite{perez_rovelli_observables} Perez and Rovelli constructed so-called W-observables, directly def\/ined on the space of dif\/feomorphism invariant spin network states\footnote{As usual in spinfoam models, the states of the theory are assumed to live on combinatorial rather embedded graphs. In this sense there is no nontrivial action of classical dif\/feomorphisms on these states. However, when passing from the quantum to the classical theory this action needs to be recovered, a point which is not clearly understood to date.}. These turn out to be related to n-point functions of group f\/ield theories. Whereas a clear connection to observables of classical gravity is absent, the same is true for n-point functions in for example quantum electrodynamics which have an inherent quantum interpretation.

An interesting, model independent analysis was presented by Giddings, Hartle and Marolf in~\cite{giddings_marolf_hartle}. Starting from classical dif\/feomorphism invariant, but nonlocal, expressions (spacetime integrals) they study in which limit these give rise to semilocal quantum operators once the theory is quantized. Interestingly they f\/ind that this is only possible for \emph{certain} observables and \emph{certain} states, related to the existence of an appropriate semiclassical limit of the theory. Furthermore, they derive  bounds on a fundamental resolution for these observables due to the interplay of quantum-ef\/fects and general background independence. In model system this gives rise to a maximum number of degrees of freedom that can be contained in any f\/inite volume.

\subsection{Dirac versus reduced phase space quantization}

Starting from a classical constraint system in general the two quantum theories obtained by Dirac quantization and reduced phase space quantization do not coincide. There are certain geometric systems with a high degree of symmetry where `quantization commutes with reduc\-tion'~\cite{guillemin_sternberg}, but in several simple model systems (see for example \cite{kunstatter, loll_reduced, reducedIII, schleich}) it was found that the two quantum theories are substantially dif\/ferent, the spectra of physical operators do not coincide. From the point of view of complete observables this can be understood as follows: following a reduced phase space quantization strategy the clock variables $T$ in the complete observables $F^\tau_{f,T}$ are replaced by real numbers (see~\cite{thiemann1} for details of this argument). Thus, the clocks $T$ are not quantized, their quantum f\/luctuations are suppressed. On the other hand, following the Dirac quantization algorithm also the clock variables $T$ are quantized, therefore the representations of complete observables within the Dirac approach will know about the quantum f\/luctuations of the clocks. In this sense, the Dirac quantization of a given constraint system can be regarded as more fundamental than the reduced phase space quantization of the same system, because all dynamical variables are treated on the same footing. In particular, when performing a reduced phase space quantization of complete observables with respect to \emph{different} clock variables $T_1 \neq T_2$ \emph{different} quantum f\/luctuations will be suppressed and the two theories will in general be inequivalent. As already stated by Thiemann in \cite{thiemann1}: ``\dots the reduced
phase space quantization of the Dirac observables can be useful only in a regime where the clocks
$T_j$ can be assumed to behave classically. This is of course not the case with respect to any choice of
clocks in extreme situations that we would like to access in quantum gravity such as at the big bang.
There we necessarily need a constraint quantization of the system.''

Subsequently, models for quantum gravity which rely on the technique of reduced phase space quantization need to be treated with care when their predictions for extreme spacetime regions are concerned because the clock variables are essentially treated as classical all the way down to the deep quantum regime. Nevertheless, such models can probably provide valuable insights into the quantum dynamics of the gravitational f\/ield in the semiclassical regime\footnote{This might sound like a severe restriction on the fundamental validity of such models. However, from an observational point of view the restriction is less severe: if quantum gravity can be experimentally tested in the far future, this will be the regime of f\/irst interest.}.

For pure general relativity or general relativity with standard matter a reduced phase space quantization seems out of reach to date, simply because the Poisson structure on the reduced phase space is way to complicated to be faithfully represented on any known Hilbert space. Thus, a Dirac quantization remains the only option as far as the full theory is concerned. Ef\/forts within the reduced phase space approach have so far been concentrated on de-parameterizable models obtained by coupling general relativity to some nonstandard matter as discussed in Section~\ref{matter_models}.

In the loop quantum gravity context the f\/irst such model was obtained by Rovelli and Smolin in~\cite{rovelli_smolin_hamiltonian} where they used the scalar f\/ield model discussed in Section~\ref{sec:scalar_fields}. Using the scalar f\/ield~$\phi$ as a clock they formally obtain the Schr\"odinger equation
\begin{gather} \label{schroedinger}
\left(i\hbar \frac{d}{d\phi} - \hat{H}_\phi\right) \psi = 0   ,
\end{gather}
where $\hat{H}_\phi$ is the quantum operator corresponding to the classical physical Hamiltonian in the scalar f\/ield reference system given by
\[
H_\phi = \int d\s \sqrt{- \sqrt{q}\fc }   .
\]
This operator is  def\/ined on the dif\/feomorphism invariant Hilbert space of (equivalence classes of) loops, therefore the contribution proportional to the dif\/feomorphism constraint in the classical expression~(\ref{scalar_field_ham}) could be omitted.

Recently Domagala, Giesel, Kaminski and Lewandowski~\cite{gravity_quantized} built upon these ideas and showed explicitly that all steps of a reduced phase space quantization can be performed for this model using the rigorous mathematical tools from loop quantum gravity. The physical Hilbert space, including a physical inner product, can be constructed explicitly\footnote{By virtue of the reduced phase space quantization the physical Hilbert space for this model has exactly the same structure as the dif\/feomorphism invariant Hilbert space for loop quantum gravity. This is essentially possible because this model does not describe pure gravity but gravity coupled to a specif\/ic scalar f\/ield. } and the physical Hamiltonian that generates time evolution in the classical reference frame def\/ined by the scalar f\/ield $\phi$ can be constructed in analogy to the Hamiltonian constraint operator in loop quantum gravity.

Also the dust model discussed in Section~\ref{sec:dust} was used as the basis for a reduced phase space quantization. As the observable algebra~(\ref{QP_Poisson}) for the model system general relativity plus pressure-less dust takes exactly the same form as the kinematical algebra for general relativity in the ADM-formalism, the physical Hilbert space for this model is obtained immediately by standard techniques and coincides with the kinematical Hilbert space of loop quantum gravity in this case. Giesel and Thiemann~\cite{AQGIV} then used loop quantum gravity methods to construct an operator version of the dust-Hamiltonian~(\ref{h_phys1}) where the main obstacle is essentially to properly implement the square root. Thus, also for this system the reduced phase space program can be completed.

Due to the complicated structure of the physical Hamiltonians in both models as operators on the physical Hilbert space the analysis could not be pushed beyond the formal level so far. Interesting questions still to be answered in this framework concern the validity of the reduced phase space approach close to gravitational singularities,  the observable ef\/fects of the dust- or scalar f\/ield matter used to de-parameterize the system and  the physical consequences of the chosen quantization of the Hamiltonian constraint. Recently progress into this direction has been made in the symmetry-reduced context, see~\cite{husain_pawlowski_2} for an application of the dust model in cosmo\-lo\-gy. The fact that all these models posses a nonvanishing physical Hamiltonian as opposed to the vanishing Hamiltonian constraint in standard loop quantum gravity raises hope that scattering theory analogous to the S-matrix approach can be developed. The key ingredient, which is still missing and unfortunately very hard to achieve, are dynamically stable coherent states for the gravitational sector needed to implement the notion of `asymptotically free states' in the scattering approach. See also~\cite{gtt2} where the dust model was analyzed using an approximation of Born--Oppenheimer type for some preliminary steps into this direction.

\subsection{Relational quantum observables}
As discussed above the de-parameterizable models are very special because they can be written in a form where at least one of the momenta appears linear in the constraints and the remaining part does not depend on the associated position variable, schematically
\[
\fc = p_t + h(q,p)  .
\]
The quantization of such models therefore can always be written in a Schr\"odinger-like form as in (\ref{schroedinger}) and the associated time evolution can be implemented \emph{unitarily}. Thus, due to the presence of a preferred reference frame (given by the dust f\/ields or the scalar f\/ields in the models discussed above) the usual incompatibility between general relativity and quantum mechanics, which manifests itself in the `problem of time', is circumvented. However, the price to pay is that (i) there now is a fundamental preferred reference frame, and (ii) the dynamical clocks used to def\/ine space and time are essentially treated as classical variables. The f\/irst point is dif\/f\/icult to reconcile with the principles of general relativity and the second with the principles of quantum mechanics. Thus, beyond the semiclassical level one needs to be careful as far as a~fully background independent theory of quantum gravity is concerned.

On the fundamental level the notion of a single universal time variable is meaningless and consequently there is no unitary time evolution. This is ref\/lected in the Dirac quantization through the requirement that physical states be annihilated by the quantum constraint operators. However, at least in a semiclassical regime, unitary evolution of quantum states with respect to an emergent (possibly observer- and state-dependent) time variable  must be possible. Because the physical Hilbert space of loop gravity is not well understood at the moment there is not much hope to clarify this issue, which is central to the interpretation of any quantum gravity theory, in the full theory. However, there has been a lot of ef\/fort to clarify the various problems that occur when one naively tries to reconcile quantum mechanics with the principle of background independence and thus constructs \emph{generally covariant quantum mechanics}, most notably by Rovelli and collaborators~\cite{hellmann_rovelli, rovelli_perez_mondragon, reisenberger_rovelli_quantum, rovelli_qm_1, rovelli_qm_2, rovelli_qm_3, rovelli_qm_wotime, rovelli_relational_qm}.

For example in \cite{rovelli_qm_wotime} a simple two-dimensional model was analyzed with a classical constraint given by
\begin{gather} \label{timeless_model}
\fc = \frac{1}{1}\big(p_1^2 + p_2^2 + q_1^2 + q_2^2\big) - M   .
\end{gather}
It is clear that this model is not de-parameterizable and does not allow a \emph{global} def\/inition of time, therefore its quantization serves as a toy model for the above mentioned issues in quantum gravity. Nevertheless, one can def\/ine time variables \emph{locally} using the framework of relational observables by considering for example $F^\tau_{q_1, q_2}$. The orbits in phase space def\/ined by~(\ref{timeless_model}) are closed, therefore $F^\tau_{q_1, q_2}$ will only be a classically well def\/ined observable for a f\/inite range of physical time $\tau$ and furthermore multi-valued. This model can trivially be (Dirac-) quantized by starting from a kinematical Hilbert space $\cH_{\rm kin} := L^2(\R^2)$ and then implementing the quantum constraint operator associated to (\ref{timeless_model}) given by
\[
\hat{\fc} = -\hbar^2 \frac{\d^2}{\d q^2_1} -\hbar^2 \frac{\d^2}{\d q^2_2} + q_1^2 + q_2^2 -(2j+1), \qquad j \in \frac{\N}{2}   .
\]
Note that the system has no quantum solutions unless for the values $M = 2j+1$. The physical Hilbert space $\cH_{\rm phys}$, def\/ined as the kernel of $\hat{\fc}$, including a well def\/ined inner product, can then explicitly be constructed.

A priori this physical Hilbert space comes without any notion of time or unitary evolution of states: the quantum observables $\hat{F}^\tau_{q_1,q_2}$ can be implemented as well def\/ined self-adjoint operators on $\cH_{\rm phys}$ but generally do not encode unitary evolution of $q_1$ with respect to $q_2$. However, it turns out that for an appropriate choice of interval $[\tau_1, \tau_2]$ \emph{and} quantum state $\psi(q_1, q_2)$ $\hat{F}^\tau_{q_1, q_2}$ is the precise analog of the Heisenberg operator $\hat{q}(t) = e^{i\hat{H}t}\hat{q} e^{-i\hat{H}t}$ and describes almost unitary evolution of $q_2$ with respect to $q_1$. Eventually the restrictions on state and time interval do not hold anymore and unitary evolution breaks down.

An interesting development concerning unitary dynamics in timeless systems was recently presented by Bojowald, H\"ohn and Tsobanjan \cite{bojowald_hoehn_tsobanjan_1, bojowald_hoehn_tsobanjan_2}: they use the ef\/fective approach towards quantum constraint systems (see for example~\cite{bojowald_effective}) where the Hilbert space picture is essentially replaced by classical equations for expectation values and moments of the quantum operators. Generally one has to consider an inf\/inite number of such momenta but for an appropriate choice of states higher order momenta are proportional to higher powers of $\hbar$ and can therefore be neglected for semiclassical considerations at a given precision. Their approach dif\/fers from the operator approach discussed above in so far as the observables, coined `fashionables' in that context, encode correlations between expectation values. Thus, they are explicitly state dependent and valid only locally. However, in contrast to other approaches a transition between dif\/ferent clocks in the quantum theory is always possible. The authors show that although a~def\/inition of unitary evolution for arbitrary long timescales is not possible in timeless systems one can choose dif\/ferent time variables for dif\/ferent regimes and then patch together these intervals of unitary evolution in a consistent way, at least as long as semiclassical behavior for at least some dynamical quantities can be assumed. For a highly excited quantum system (such as quantum gravity close to the big bang) this description breaks down and nicely illustrates the intuition that quantum mechanical time evolution cannot be more than a semiclassical concept.

Recently Gambini, Pullin and collaborators have argued that the fundamental interpretation of quantum mechanics needs to be changed in order to accommodate for background independent quantum gravity where a strict separation of nondynamical measurement devices from the quantum dynamics of the system is not possible. See for example  \cite{ montevideo1, montevideo2, montevideo3, montevideo4} for the framework coined the `Montevideo interpretation' of quantum mechanics.

\subsection{Spectra of physical operators in loop quantum gravity?}

One of the most interesting features of loop quantum gravity is that geometrical operators, such as the quantum versions of areas and volumina, have discrete spectra. Thus, the intuitive idea that smooth spacetime is replaced by a fundamentally discrete entity at the Planck scale is realized in a precise way in that formalism. Moreover, this Planck scale discreteness is not put in by hand but emerges naturally in the background independent quantization used in loop quantum gravity.

One of the open questions, still under debate, is whether the discreteness of these spectra is merely a kinematical artefact or also holds on the level of the physical Hilbert space. Area- and volume-operators are def\/ined on the \emph{kinematical} (or, depending on the interpretation, spatial dif\/feomorphism invariant) Hilbert space, and a priori one would have to construct their gauge-invariant extensions to construct geometrical operators acting on the \emph{physical} Hilbert space. However, this Hilbert space is under poor control so far. Thus, a detailed analysis as for the  toy model presented above is not available.

In \cite{dittrich_thiemann_spectra} Dittrich and Thiemann analyzed several f\/inite dimensional models and showed that a priori an operator that has discrete spectrum on the kinematical Hilbert space \emph{may} have continuous spectrum when extended in a gauge invariant way to the physical Hilbert space. Thus, they conclude that a more detailed analysis within the context of full loop quantum gravity is needed to decide whether physical areas and volumina are discrete or not.

In \cite{rovelli_spectra} Rovelli argued that the models considered in \cite{dittrich_thiemann_spectra} show some pathological behavior not present in full loop quantum gravity related to a subtlety of compact versus noncompact orbits of the corresponding classical observables in phase space. Furthermore, a number of general physical arguments is provided in favor of the hypothesis that geometrical spectra remain discrete on the physical Hilbert space. One argument, which f\/its within the recent attempts to construct a~reduced phase space quantization of gravity discussed above is the following: in general relativity the evolution of coordinate-dependent quantities is under-determined. This can be interpreted in two dif\/ferent ways. Either, one takes the point of view that only coordinate-independent quantities can possibly have a physical meaning. This is essentially the point of view taken when constructing complete observables for pure general relativity or general relativity with standard matter and subsequently applying a Dirac quantization. Or, one takes the point of view that the coordinates $x^\mu$ describe \emph{physical} positions with respect to some physical dynamical reference frame which, however, one does not care describing. The under-determinacy of the evolution then ref\/lects the fact that the dynamical evolution of the reference frame is not taken into account\footnote{See also \cite{frank_observables} where the construction of complete observables for extended systems is studied and a similar conclusion is reached.}. Thus the seemingly gauge-dependent quantities are truly gauge-invariant observables of the extended system where the clock f\/ields have been gauge f\/ixed to some physical value.

This idea is realized in a precise way in the reduced phase space quantizations for gravity \cite{gravity_quantized,AQGIV, husain_pawlowski_1} where discrete spectra for geometrical operators on the physical Hilbert space are found. However, as expressed above there are certain doubts concerning the validity of the reduced phase space quantization in the deep quantum regime and, consequently, whether physical predictions of these models at the Planck scale must be corrected or not.

Furthermore, one can argue that the measurable discreteness of the spectrum of some operator might be a \emph{kinematical} property and not a dynamical one after all: consider a classical system where the motion of a free particle is constrained to a circle, thus its position is $q \in [0,2\pi]$. Upon quantization it turns out that the canonical momentum $p$ has discrete spectrum. One easy way to see this is the following: position and momentum coordinates are related by a Fourier transform. The Fourier transform of a function on a compact space has support only on a discrete space. Now, add a nontrivial smooth bounded potential $V(q)$ to the Hamiltonian describing that system. For generic potentials the quantum dynamics can no longer be solved anymore, that is the `physical' Hilbert space can no longer be constructed. Nevertheless, the spectrum of the momentum operator is discrete, as is conf\/irmed by numerous experimental results. Thus, from this perspective one would conclude that discreteness of spectra is a kinematical property. In general covariant theories this argument is then further complicated by the interlink between dynamics and constraints.

To summarize, it is still an open question whether the \emph{physical} spectra of geometrical ope\-ra\-tors are discrete or not as there are good arguments for both points of view. Much of this confusion is linked to the lack of experience with background independent quantum mechanics, where the principles of standard quantum mechanics will have to be generalized in one way or the other. Unfortunately an answer in the full theory seems out of reach to date, among other things due to mathematical complexity. Symmetry reduced models, such as loop quantum cosmology~\cite{ashtekar_singh_report, bojowald_review}, might provide a good test f\/ield to study conceptual questions.

\section{Discussion and open questions}

In this article we presented an overview on the topic of observables in gravity. As far as classical gravity is concerned there has been a lot of progress in recent years, at least on the conceptual side. The peculiar case of general relativity, where the gauge group is the dif\/feomorphism group $\Diff(M)$ and therefore is strongly related to the choice of observer, required the developments of new tools and techniques to cope with mathematical and conceptual dif\/f\/iculties. The method of \emph{complete observables} encodes in a mathematically precise way the intuitive idea that relations between dynamical quantities are all that matters in a background independent theory. This led to a better understanding of the observable structure of general relativity, which is thought to be an important ingredient for any viable theory of quantum gravity.

\looseness=-1
However, due to the complicated structure of the gauge group $\Diff(M)$ analytic results for full general relativity coupled to realistic matter f\/ields are to a large extend still missing. The initial question how local gravitational excitations, such as gravitational waves, can be described analy\-ti\-cal\-ly in a fully gauge- (and thus observer-) invariant way is not answered yet. This is not to say that the problem is poorly understood: there exists a formalism which provides a~precise recipe to compute gauge invariant observables for general relativity which is conceptually very well understood. However, calculations are typically so hard that they cannot be solved analytically.

This is a common feature of practically any realistic theory and now one should follow the scheme which has led to success in many cases: stop bothering about formal details, develop approximation methods and extract physical predictions from this formalism.

We described f\/irst attempts into this directions in this article, namely the perturbative expansion of complete observables around a f\/ixed spacetime. Within this formalism one can see, among other things, how the observables of linearized gravity (such as gravitational waves) emerge as f\/irst order approximations to observables of the full theory. Thus, familiar `low curvature'-physics can be seen to emerge from a fully background independent context. However, as it stands the perturbative formalism is much to complicated to ef\/f\/iciently address questions of true physical interest and needs to be substantially enhanced to be useful for practical calculations. One big unknown, which has a lot of impact on the tractability of the computations, is the choice of clock variables and thus dynamical reference frame.

\looseness=-1
We discussed a class of model systems with a specif\/ic choice of matter f\/ields that leads to a complete de-parameterization when coupled to general relativity. This allows the \emph{analytic} computation of gauge invariant observables for a generally covariant theory and is therefore of outstanding mathematical interest. Unfortunately these matter models seem rather artif\/icial and it is unlikely that they can stand the comparison with phenomenology in gravitational conf\/igurations of any kind: the implicit assumption within these models is that there is one single set of clock variables which provides a good reference frame for any possible physical situation.

The latter assumption in a sense contradicts the spirit of general relativity: there might be no preferred reference frame at all and good clock variables have to be chosen case by case, depending on which gravitational conf\/iguration one wishes to describe. However, within the framework of complete observables this does a priori not pose any problems. It will be interesting to see whether the analytic developments from the de-parameterized models can be lifted to the general case.

\looseness=-1
We also reviewed the status of gauge invariant observables in loop quantum gravity. As the physical Hilbert space is so far under poor control the detailed construction of complete quantum observables for gravity has not been possible so far. However, the peculiarities that one has to face when quantizing systems without a global time variable are under investigation and we have discussed a framework that, in the spirit of complete observables, allows to derive dif\/ferent domains of semiunitary evolution from timeless quantum systems. This works as long as there are at least some variables which behave semiclassically and coincides with the expectation that unitary time evolution might become meaningless in the deep quantum regime close to the big bang.

These obstructions can be side-stepped by quantizing the de-parameterizable models in a~reduced phase space quantization: as there exists a preferred reference frame and the clock variables are essentially treated as classical all the way down to the deep quantum regime time evolution in that reference frame is implemented unitarily. To date it is unclear how trustworthy the reduced phase space quantization is when physics close to gravitational singularities is concerned, because the quantum f\/luctuations of the clock variable seem to be artif\/icially suppressed. From this point of view it seems to be important to compare the physical predictions obtained within models based on dif\/ferent reduced phase spaces. Thus, one can check whether they describe the same quantum theory, at least in the semiclassical sector.

For the full theory, where a preferred notion of time in the quantum regime is absent, it will be important to further push the development of approximation tools. We mentioned the mode expansion and the use of ef\/fective equations that can convincingly be adapted to timeless quantum theories. We hope that such methods can be generalized to shed further light on the question of quantum observables in full loop quantum gravity.

\subsection*{Acknowledgements}

I want to thank Carlo Rovelli for comments on a previous draft of this article and for helpful suggestions. Discussions with Bianca Dittrich and Wojciech Kaminski are gratefully appreciated. The author is partially supported by the ANR ``Programme Blanc" grants LQG-09.
\pdfbookmark[1]{References}{ref}
\LastPageEnding

\end{document}